\documentclass[12pt,draftclsnofoot,onecolumn]{IEEEtran}

\IEEEoverridecommandlockouts
\usepackage{indentfirst,flushend}
\usepackage{amssymb,bm,mathrsfs,times,amscd,amsmath,amsthm,amsfonts,bbm}
\usepackage{latexsym}
\usepackage{dsfont,diagbox}
\usepackage{graphicx}
\usepackage{subfigure}
\usepackage{color}
\usepackage{cases}

\usepackage{multirow}
\usepackage[sort]{cite}
\usepackage{multicol}
\usepackage[nooneline,flushleft]{caption2}
\usepackage{clrscode}
\usepackage{algorithm}
\usepackage{psfrag,stfloats,nicefrac}
\usepackage{algorithmic}
\usepackage{supertabular}
\usepackage{makecell}
\allowdisplaybreaks[4]

\begin{document}

\title{Cloud-Edge Coordinated Processing: Low-Latency Multicasting Transmission}
\author{Shiwen~He,~\IEEEmembership{Member,~IEEE}, Ju Ren,~\IEEEmembership{Member,~IEEE},\\~Jiaheng Wang,~\IEEEmembership{Senior Member,~IEEE},~Yongming Huang,~\IEEEmembership{Senior Member,~IEEE}, \\Yaoxue~Zhang,~\IEEEmembership{Senior Member,~IEEE}, Weihua Zhuang,~\IEEEmembership{Fellow, IEEE}, and\\ Sherman (Xuemin) Shen,~\IEEEmembership{Fellow, IEEE}
\thanks{S. He, J. Ren, and Y. Zhang are with the School of Computer Science and Engineering, Central South University, Changsha 410083, China. They are also with the School of information Technology, Jiangxi University Of Finance and Economics, West Yuping Road, Nanchang 330032, China. (email: \{shiwen.he.hn, renju, zyx\}@csu.edu.cn). }
\thanks{J. Wang and Y. Huang are with the National Mobile Communications Research Laboratory, School of Information Science and Engineering, Southeast University, Nanjing 210096, China. (email: \{jhwang, huangym\}@seu.edu.cn). }
\thanks{S. Shen and W. Zhuang are with the Department of Electrical and Computer Engineering, University of Waterloo, ON, Canada N2L 3G1. (email: \{sshen, wzhuang\}@uwaterloo.ca). }
}

\maketitle
\vspace{-.6 in}

\begin{abstract}
Recently, edge caching and multicasting arise as two promising technologies to support high-data-rate and low-latency delivery in wireless communication networks. In this paper, we design three transmission schemes aiming to minimize the delivery latency for cache-enabled multigroup multicasting networks. In particular, full caching bulk transmission scheme is first designed as a performance benchmark for the ideal situation where the caching capability of each enhanced remote radio head (eRRH) is sufficient large to cache all files. For the practical situation where the caching capability of each eRRH is limited, we further design two transmission schemes, namely partial caching bulk transmission (PCBT) and partial caching pipelined transmission (PCPT) schemes. In the PCBT scheme, eRRHs first fetch the uncached requested files from the baseband unit (BBU) and then all requested files are simultaneously transmitted to the users. In the PCPT scheme, eRRHs first transmit the cached requested files while fetching the uncached requested files from the BBU. Then, the remaining cached requested files and fetched uncached requested files are simultaneously transmitted to the users. The design goal of the three transmission schemes is to minimize the delivery latency, subject to some practical constraints. Efficient algorithms are developed for the low-latency cloud-edge coordinated transmission strategies. Numerical results are provided to evaluate the performance of the proposed transmission schemes and show that the PCPT scheme outperforms the PCBT scheme in terms of the delivery latency criterion.
\end{abstract}

\section*{\sc \uppercase\expandafter{\romannumeral1}. Introduction}

Driven by the visions of ultra-high-definition video, intelligent driving, and Internet of Things, high-data-rate and low-latency delivery become two key performance indicators of future wireless communication networks~\cite{AccessGupa2015}. The vast resources available in the cloud radio access server can be leveraged to deliver elastic computing power and storage to support resource-constrained end-user devices~\cite{CSTCheck2015}. However, it is not suitable for a large set of cloud-based applications such as the delay-sensitive ones, since end devices in general are far away from the central cloud server, i.e., data center~\cite{Ju2017,Deyu2019}. To overcome these drawbacks, caching popular content at the network edge during the off-peak period is proved to be a powerful technique to realize low-latency delivery for some specific applications, such as real-time online gaming, virtual reality, and ultra-high-definition video streaming, in next-generation communication systems~\cite{Shen2018,ATNPedars2016,TSTMour2018}. Consequently, an evolved network architecture, labeled as cache-enabled radio access networks (RANs), has emerged to satisfy the demands of ultra-low-latency delivery by migrating the computing and caching functions from the cloud to the network edge~\cite{CommBast2014,TSTMour2018,NetvPeng2016}. In the cache-enabled RANs, the cache-enabled radio access nodes named as enhanced remote radio heads (eRRHs) have the ability to enable processing at the network edge and to cache files at its local cache~\cite{NetvPeng2016,NetvShih2017,TITSeng2017}.

\subsection*{A. Related Works}

The main motivation of caching frequently requested content at the network edge is to reduce the burden on the fronthaul links and the delivery latency. Existing studies on the cache-enabled RANs are mainly on two-fold, i.e., the pre-fetching phase and the delivery phase. The pre-fetching phase studies focus on caching strategies, while accounting for the caching capacity of eRRHs, popularity of contents and user distribution~\cite{TWCLiu2017,CLZheng2017,TWCZhu2018,TWCKwak2018,TCOMCui2018,TCOMWen2018,TWCZheng2018}. The delivery phase deals with the requested data transmission for different performance criteria~\cite{TWCPark2016,TWCHe2019,TWCTao2016,JSACLiu2016,TWCXiang2018,TWCHajri2018,TCOMHe2018}.

\emph{1) Optimization of content placement:} Content placement with a finite cache size is the key issue in caching design, since unplanned caching at the network edge will result in more inter-cell interference or delivery latency. Therefore, how to effectively cache the popular content at the network edge has attracted extensive attention from both academia and industry. The cache placement problem in cache-enabled RANs is investigated while accounting for the flexible physical-layer transmission and diverse content preferences of different users~\cite{TWCLiu2017}. Hybrid caching together with relay clustering is studied to strike a balance between the signal cooperation gain achieved by caching the most popular contents and the largest content diversity gain by caching different contents~\cite{CLZheng2017}. In~\cite{TWCZhu2018,TWCKwak2018,TCOMCui2018,TCOMWen2018,TWCZheng2018}, an edge caching strategy is investigated for cache-enabled coordinated heterogeneous networks. Probabilistic content placement is designed to maximize the performance of content delivery for cache-enabled multi-antenna dense small cell network~\cite{TWCZhu2018}. Dynamic content caching is studied via stochastic optimization for a hierarchical caching system consisting of original content servers, central cloud units and base stations (BSs)~\cite{TWCKwak2018}. The successful transmission probabilities are analyzed for a two-tier large-scale cache-enabled wireless multicasting network~\cite{TCOMCui2018,TCOMWen2018}. Proactive caching strategies are proposed to reduce the backhaul transmission for large-scale mobile edge networks~\cite{TWCZheng2018}. Though efficient caching at the network edge can effectively reduce the burden on the fronthaul links, how to effectively design a transmission strategy is key problem, especially for content-centric ultra-dense massive-access networks.

\emph{2) Optimization of transmission strategy:} How to timely transmit the cached and uncached requested files to users is another key problem for cache-enabled coordinated RANs~\cite{TWCPark2016,TWCHe2019,TCOMHe2018,TWCTao2016,JSACLiu2016,TWCHajri2018,TWCXiang2018}. Joint optimization of cloud-edge coordinated precoding using different pre-fetching strategies is investigated respectively for cache-enabled sub-6 GHz and millimeter-wave multi-antennas multiuser RANs in~\cite{TWCPark2016,TWCHe2019}. He \emph{et al.} propose a two-phase transmission scheme to reduce both burden on the fronthaul links and delivery latency with a fixed delay caused by fronthaul links for cache-enabled RANs~\cite{TCOMHe2018}. Tao \emph{et al.} investigate the joint design of multicast beamforming and eRRH clustering for the delivery phase with fixed pre-fetching to minimize the compound cost including the transmit power and fronthaul cost, subject to predefined delivery rate requirements~\cite{TWCTao2016}. Research on the energy efficiency of cache-enabled RANs shows that caching at the BSs can improve the network energy efficiency when power efficient cache hardware is used~\cite{JSACLiu2016,TWCHajri2018}. Studies on cache-enabled physical layer security have shown that caching can reduce the burden on fronthaul links, introduce additional secure degrees of freedom, and enable power-efficient communication~\cite{TWCXiang2018}. However, maximizing the (minimum) delivery rate or minimizing the compound cost function may not necessarily minimize the delivery latency, especially when partial requested contents are not cached at the network edge.

\subsection*{B. Contributions and Organization}

In general, the delivery latency is incurred at least by the propagation of fronthaul links, signal processing at the BBU, and signal transmission for wireless communication systems. Furthermore, the limited capacity of fronthaul links is a key factor in determining the delivery latency and is the key motivation for mitigating the cache and baseband signal processing to the network edge. However, to the best of the authors' knowledge, how to effectively exploit the cache and baseband signal processing at the network edge to minimize the delivery latency is an open problem for cache-enabled multi-antennas multigroup multicasting RANs with limited capacities of fronthaul links. In this paper, we study the minimization of delivery latency of three different transmission schemes, accounting for the delay caused by fetching the uncached requested files from the BBU and the signal processing at the BBU for cache-enabled multi-antennas multigroup multicasting RANs. The main contributions of this paper can be summarized as follows:
\begin{itemize}
\item When the caching capability of each eRRH is sufficient large to cache all files, a full caching bulk transmission (FCBT) scheme is proposed as a performance benchmark for minimizing the delivery latency;

\item In practice, only a part of files is cached at network edge due to the limited caching capability of each eRRH. For this case, we first present a partial caching bulk transmission (PCBT) scheme that transmits simultaneously all requested files to the users and then a novel partial caching pipelined transmission (PCPT) scheme to further reduce the delivery latency;

\item Three optimization problems are formulated to minimize the delivery latency that includes the delay caused by fetching the uncached files from the BBU, signal processing at the BBU, and transmitting all requested files to the users, subject to constraints on the fronthaul capacity, per-eRRH transmit power constraint, and file size;

\item An efficient algorithm that is proved to converge to a Karush-Kuhn-Tucker (KKT) solution is designed to solve each of optimization problems, respectively;

\item Numerical results are provided to validate the effectiveness of the proposed methods. Compared to the other non-FCBT transmission schemes, the PCPT scheme achieves obvious performance improvement in terms of delivery latency.
\end{itemize}

The remainder of this paper is organized as follows. The system model is described in Section \uppercase\expandafter{\romannumeral2}. In Section \uppercase\expandafter{\romannumeral3}, three transmission schemes are formulated. Design of optimization algorithms are investigated in Section \uppercase\expandafter{\romannumeral4}. Numerical results are presented in Section~\uppercase\expandafter{\romannumeral5} and conclusions are drawn in Section~\uppercase\expandafter{\romannumeral6}.

\textbf{Notations}: Bold lower case and upper case letters represent column vectors and matrices, respectively; $\mathrm{diag}\left(\mathbf{a}\right)$ is a diagonal matrix whose diagonal elements  are the elements of vector $\mathbf{a}$; $\mathbf{0}_{N\times N}$ and $\mathbf{I}_{N\times N}$ denote the $N\times N$ zero and identity matrices, respectively; $\mathrm{tr}\left(\cdot\right)$, $\|\cdot\|_{2}$,  and $\|\cdot\|_{\mathrm{F}}$ denote the trace, the Euclidean norm, and the Frobenius norm, respectively. The circularly symmetric complex Gaussian distribution with mean $\mathbf{u}$ and covariance matrix $\mathbf{R}$ is denotes by $\mathcal{CN}\left(\mathbf{u}, \mathbf{R}\right)$; $\mathbf{A}\succeq \bm{0}$ is a positive semidefinite matrix; $\left[\mathbf{A}\right]_{\left(m,n\right)}$ represents the element in row $m$ and column $n$ of matrix $\mathbf{A}$, and $\mathrm{vec}\left(\mathbf{A}\right)$ denotes the column vector obtained by stacking the columns of matrix $\mathbf{A}$ on top of one another. Superscripts $\left(\cdot\right)^{T}$, $\left(\cdot\right)^{*}$, and $\left(\cdot\right)^{H}$ represent transpose, conjugate, and conjugate transpose operators, respectively. For set $\mathcal{A}$, $\left|\mathcal{A}\right|$ denotes the cardinality of the set, while for complex number $x$, $\left|x\right|$ denotes the magnitude value of $x$; $\mathbb{R}$ and $\mathbb{C}$ are the fields of real and complex numbers, respectively. Function $\lfloor x\rfloor$ rounds $x$ to the nearest integer not larger than $x$; $\overline{a}$ denotes the complement $1-a$ of a binary variable $a\in\left\{0,1\right\}$; and $\ln\left(\cdot\right)$ is the logarithm with base $e$. The circularly symmetric complex Gaussian distribution with mean $\mathbf{\mu}$ and covariance matrix $\mathbf{R}$ is denoted by $\mathcal{CN}\left(\mathbf{\mu},\mathbf{R}\right)$. The symbols used frequently in this paper are summarized in Table~\ref{SymbolSummarized}.
\begin{table*}[htbp]
	\setlength{\abovecaptionskip}{0pt}
	\setlength{\belowcaptionskip}{5pt}
\renewcommand{\captionfont}{\footnotesize}
\renewcommand*\captionlabeldelim{.}
	\captionstyle{flushleft}
	\onelinecaptionstrue
	\centering
	\caption{List of important mathematical symbols.}
	\begin{tabular}{|c|l|c|l|}
		\hline
		Symbol&\makecell[c]{Meaning}&Symbol&\makecell[c]{Meaning}\\
		\hline
		\hline
		$K_{\mathrm{U}}$, $K_{\mathrm{R}}$ &Numbers of users and eRRHs&$\mathcal{K}_{\mathrm{U}}$, $\mathcal{K}_{\mathrm{R}}$ &Sets of users and eRRHs\\
		\hline
        $N_{\mathrm{t}}$ &Numbers of antennas equipped at each eRRH&		
		$P_{i}$ &Maximum transmit power of eRRH $i$\\
		\hline
		$C_{i}$ &Capacity of fronthaul link to eRRH $i$&
		$B_{i}$ &Normalized cache size of eRRH $i$\\
        \hline
        $G$ & Number of multicast groups & $G_{g}$ &$g$-th group\\
		\hline
		$F$ &Number of files in the library&
		$\mathcal{F}$, $\mathcal{F}_{\mathrm{req}}$ &Sets of all files and all requested files\\
		\hline
        $S$ &Normalized size of files&
		$f_{g}$ &Index of file requested by the $g$-th group\\
		\hline
		$c_{f,i}$ &Binary caching variable of file $f$ at eRRH $i$&
		$\overline{c}_{f,i}$ &Complement of $\overline{c}_{f,i}$\\
		\hline
        $\mathbf{y}_{k}$ &Signal received by user $k$&
		$\mathbf{h}_{k,i}$ &Channel matrix from eRRH $i$ to user $k$\\
		\hline
        $\sigma_{k}^{2}$ & Noise variance at user $k$&
		$\mathcal{V}$ &Set of beamforming vector $\mathbf{v}_{g,i}$\\
		\hline
		$\mathbf{\Omega}_{i}$ &Quantization noise covariance matrix of eRRH $i$&
		$\mathcal{O}$ &Set of quantization noise covariance matrices\\
		\hline
        $\gamma_{k,i}$ &\multicolumn{3}{l|}{SINR of user $k$ for full caching case or partial caching case}\\
		\hline
		$r_{k,i}$ &\multicolumn{3}{l|}{Achievable rate of user $k$ for full caching case or partial caching case, $i=1,2$}\\
		\hline
        $\mathbf{w}_{g,i}$ &\multicolumn{3}{l|}{Beamforming vector for basedband signal $\mathbf{s}_{f_{g}}$ at eRRH $i$ for full caching cse}\\
		\hline
        $\mathbf{u}_{g,i}$ &\multicolumn{3}{l|}{Beamforming vector for cached basedband signal $\mathbf{s}_{f_{g}}$ at eRRH $i$ for partial caching case}\\
		\hline
        $\mathbf{v}_{g,i}$ &\multicolumn{3}{l|}{Beamforming vector for uncached basedband signal $\mathbf{s}_{f_{g}}$ at eRRH $i$ for partial caching case}\\
		\hline
		$\mathbf{h}_{k}$, $\mathbf{w}_{g}$ &\multicolumn{3}{l|}{$\mathbf{h}_{k}=\left[\mathbf{h}_{k,1}^{H},\cdots, \mathbf{h}_{k,K_{\mathrm{R}}}^{H}\right]^{H}$, $\mathbf{w}_{g}=\left[c_{f_{g},1}\mathbf{w}_{g,1}^{H},\cdots,c_{f_{g},K_{\mathrm{R}}}
\mathbf{w}_{g,K_{\mathrm{R}}}^{H}\right]^{H}$}\\
		\hline
		$\overline{\mathbf{w}}_{k}$, $\mathbf{u}_{g}$, $\mathbf{u}_{g}$ &\multicolumn{3}{l|}{$\overline{\mathbf{w}}_{g}=\mathbf{u}_{g}+\mathbf{v}_{g}$, $\mathbf{u}_{g}=\left[c_{f_{g},1}\mathbf{u}_{g,1}^{H},\cdots,c_{f_{g},K_{\mathrm{R}}}\mathbf{u}_{g,K_{\mathrm{R}}}^{H}
\right]^{H}$, $\mathbf{v}_{g}=\left[\overline{c}_{f_{g},1}\mathbf{v}_{g,1}^{H},
\cdots,\overline{c}_{f_{g},K_{\mathrm{R}}}\mathbf{v}_{g,K_{\mathrm{R}}}^{H}\right]^{H}$}\\
		\hline
	\end{tabular}
	\label{SymbolSummarized}
\end{table*}

\section*{\sc \uppercase\expandafter{\romannumeral2}. System Model}

\begin{figure}[t]
\renewcommand{\captionfont}{\footnotesize}
\renewcommand*\captionlabeldelim{.}
	\centering
	\captionstyle{flushleft}
	\onelinecaptionstrue
	\includegraphics[width=0.6\columnwidth,keepaspectratio]{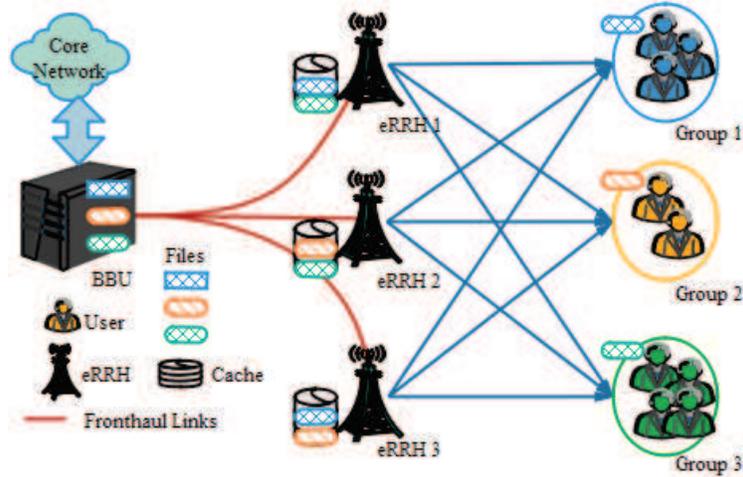}\\
	\caption{Illustration of downlink transmission of a cache-enabled multigroup multicasting RAN.}
	\label{EquivalentSystemModel}
\end{figure}

Consider the downlink transmission of a cache-enabled multigroup multicasting RAN, as illustrated in Fig.~\ref{EquivalentSystemModel}, comprising one baseband unit (BBU), $K_{\mathrm{R}}$ eRRHs, and $K_{\mathrm{U}}$ single-antenna users.
In the system, eRRH $i\in\mathcal{K}_{\mathrm{R}}=\left\{1,\cdots,K_{\mathrm{R}}\right\}$ is equipped with a cache that can store $B_{i}$ nats, where $B_{i}$ is the normalized cache size;  eRRH $i$ is equipped with $N_{\mathrm{t}}$ antennas and connected to the BBU through an error-free fronthaul link with normalized capacity $C_{i}$ nats/Hz/s. The user set $\mathcal{K}_{\mathrm{U}}=\left\{1,\cdots,K_{\mathrm{U}}\right\}$ is partitioned into $G$ multicast groups, denoted by $\mathcal{G}_{1}$, $\cdots$, $\mathcal{G}_{G}$. Each user, $k\in\mathcal{K}_{\mathrm{U}}$, independently requests only a single file in a given transmission interval, i.e., each user belongs to at most one multicast group.

Without loss of generality, we assume that all files in library $\mathcal{F}=\left\{1,\cdots,F\right\}$ at the BBU have the same size of $S$ nats/Hz, where $S$ is the normalized file size. The assumption of equal file sizes is standard and reasonable in that the most frequently requested and cached files by users are chunks of videos, e.g., fragments of a given duration, which are often segmented with the same length~\cite{ATNPedars2016}. In general, eRRH $i\in\mathcal{K}_{\mathrm{R}}$ selectively pre-fetches some popular files from library $\mathcal{F}$ to its local cache during the off-peak period, according to the content popularity and predefined caching strategies~\cite{TWCPark2016,TWCTao2016}. The cache status of file $f\in\mathcal{F}$ can be modeled as a binary variable $c_{f,i}$, $f\in\mathcal{F}$, $i\in\mathcal{K}_{\mathrm{R}}$, given by
\begin{equation}\label{DelayAware01}
c_{f,i}=
\begin{cases}
1, ~~\text{if file}~f~\text{is cached by eRRH}~i, \\
0 , ~~\text{otherwise} .
\end{cases}
\end{equation}
The complement of $c_{f,i}$ is denoted as $\overline{c}_{f,i}=1-c_{f,i}$. In this work, we focus on transmission strategies given cache state information, i.e., $c_{f,i}$, $f\in\mathcal{F}$, $i\in\mathcal{K}_{\mathrm{R}}$. Let $\mathcal{F}_{\mathrm{req}}=\cup_{g\in\mathcal{G}}\left\{f_{g}\right\}\subseteq\mathcal{F}$ be the index set of requested files of all user groups, where $f_{g}$ is the index of the requested file by the users in group $\mathcal{G}_{g}$. We denote the respective group index of user $k$ by a positive integer $g_{k}$.

\section*{\sc \uppercase\expandafter{\romannumeral3}. Design of Transmission Schemes}

In this section, we consider three transmission schemes according to the caching capabilities of eRRHs and then formulate the corresponding optimization problems for cache-enabled multigroup multicasting RANs. In particular, the design objective is to minimize the delivery latency subject to the constraints on the fronthaul links, eRRH transmit power, and file size.

\subsection*{A. Full Caching Bulk Transmission (FCBT) Scheme}

In this subsection, we assume that all files are cached at the local cache, i.e., $c_{f,i}=1$, $\forall i\in\mathcal{K}_{\mathrm{R}}$, $\forall f\in\mathcal{F}$. Consequently, all requested files can be directly retrieved from the local caches of eRRHs and transmitted to the users\footnote{The reason of considering full caching is to provide a benchmark for performance comparison. In practical communication networks, eRRH cannot cache all requested files even if it has sufficient large caching capacity due to the diversity of files, massive users, and mobility of users.}. We refer to this transmission scheme as full caching bulk transmission (FCBT) scheme, which is similar with the coordinated multiple points (CoMP) mechanism in wireless communication systems~\cite{TSPZhuang2014}.

In the FCBT scheme, the users first send the requirement of files to the eRRHs and then the eRRHs coordinately transmit the requested files to the users. Consequently, the received baseband signal at user $k$ can be expressed as
\begin{equation}\label{DelayAware02}
y_{k,1}=\sum\limits_{g\in\mathcal{G}}
\mathbf{h}_{k}^{H}\mathbf{w}_{g}s_{f_{g}}+n_{k}
\end{equation}
where $\mathbf{w}_{g}=\left[c_{f_{g},1}\mathbf{w}_{g,1}^{H},\cdots,c_{f_{g},K_{\mathrm{R}}}
\mathbf{w}_{g,K_{\mathrm{R}}}^{H}\right]^{H}$, with $\mathbf{w}_{g,i}\in\mathbb{C}^{N_{\mathrm{t}}\times 1}$ being the beamforming vector for the users in $\mathcal{G}_{g}$ at eRRH $i$, $\mathbf{h}_{k}=\left[\mathbf{h}_{k,1}^{H},\cdots,\mathbf{h}_{k,K_{\mathrm{R}}}^{H}\right]^{H}$ with
$\mathbf{h}_{k,i}\in\mathbb{C}^{N_{\mathrm{t}}\times 1}$ denoting the channel coefficient between user $k$ and eRRH $i$, $s_{f_{g}}$ represents the baseband signal of requested file $f_{g}$ for the users in $\mathcal{G}_{g}$, and $n_{k}$ denotes the additive white Gaussian noise with $\mathcal{CN}\left(0,\sigma_{k}^{2}\right)$. Thus, the signal-to-interference-plus-noise ratio (SINR) of user $k$ is calculated as
\begin{equation}\label{DelayAware03}
\gamma_{k,1}=\frac{\left|\mathbf{h}_{k}^{H}\mathbf{w}_{g_{k}}\right|^{2}}
{\sum\limits_{g\in\mathcal{G}\setminus\left\{g_{k}\right\}}
\left|\mathbf{h}_{k}^{H}\mathbf{w}_{g}\right|^{2}+\sigma_{k}^{2}}.
\end{equation}
Based on Shannon capacity, the corresponding achievable rate on unit bandwidth in one second of user $k$ is given by $R_{k,1}=\ln\left(1+\gamma_{k,1}\right)$ in unit of nats/Hz/s. Let $r_{g,1}$ in unit of nats/Hz/s be the delivery rate of group $g$ for the cached files transmission. We consider minimizing the delivery latency and meanwhile providing fairness for all multicast groups. To achieve this two-fold goal, the design problem is formulated as
\begin{subequations}\label{DelayAware09}
\begin{align}
&\min\max_{g\in\mathcal{G}}\frac{S}
{\mathrm{r}_{g,1}}\label{DelayAware09a}\\
\mathrm{s.t.}~&\mathrm{r}_{g,1}\leq\mathrm{R}_{k,1}, \forall k\in\mathcal{G}_{g}, \forall g\in\mathcal{G}\label{DelayAware09b}\\
&\sum\limits_{g\in\mathcal{G}}\left\|c_{f_{g},i}\mathbf{w}_{g,i}\right\|^{2}\leq P_{i}, \forall i\in\mathcal{K}_{\mathrm{R}}\label{DelayAware09c}
\end{align}
\end{subequations}
where the optimization variables are $\mathbf{w}_{g}$ and $\mathrm{r}_{g,1}$, $\forall g\in\mathcal{G}$.  Constraints~\eqref{DelayAware09b} means that the delivery rate of the file requested by user $k$ is constrained by the achievable rate. Constraints~\eqref{DelayAware09c} is the power constraint per eRRH. The goal of problem~\eqref{DelayAware09} is to minimize the maximum group delivery latency, as the accomplishment of the transmission of requested files is determined by the worst group for the whole data transmission.

\subsection*{B. Partial Caching Bulk Transmission Scheme}
\begin{figure}[t]
\renewcommand{\captionfont}{\footnotesize}
\renewcommand*\captionlabeldelim{.}
	\centering
	\captionstyle{flushleft}
	\onelinecaptionstrue
	\includegraphics[width=1\columnwidth,keepaspectratio]{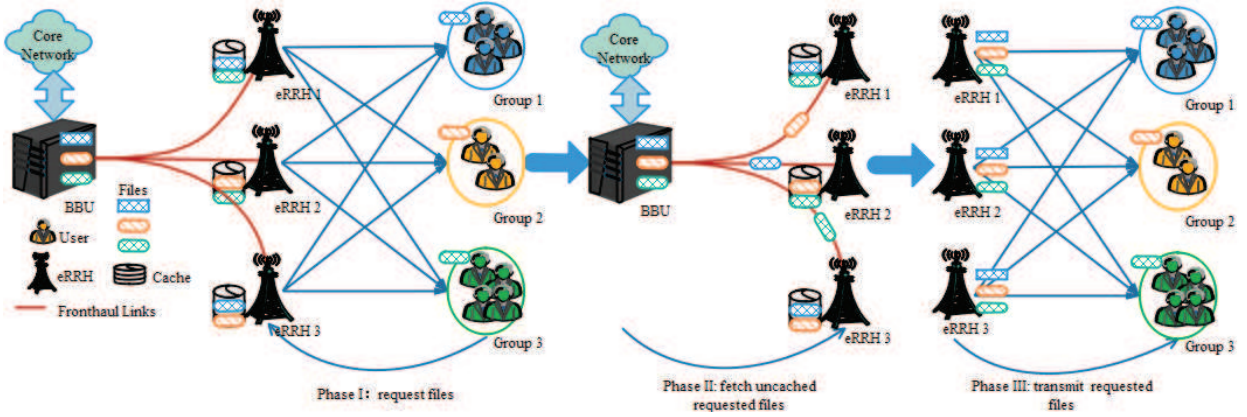}\\
	\caption{Flowchart of partial caching buck transmission scheme.}
	\label{SinglePhaseFlowChart}
\end{figure}

In practice, each eRRH can only cache a fractional of files at its local cache due to the limited caching capabilities of eRRHs. As a result, some requested files may not be cached at the cache of the network edge. In this subsection, we explore the problem of minimizing the delivery latency for the case that only a fractional of requested files are cached at the local cache of eRRHs. In this case, a common transmission scheme contains three phases~\cite{TWCPark2016}, as illustrated in Fig.~\ref{SinglePhaseFlowChart}, which is termed as partial caching bulk transmission (PCBT) scheme. In particular, in phase \uppercase\expandafter{\romannumeral1}, the users first send the requirement of files to the eRRHs. Because only partial required files are cached at the local cache, the eRRHs fetch the uncached requested files from the BBU in phase \uppercase\expandafter{\romannumeral2}. After obtaining the uncached requested files, in phase \uppercase\expandafter{\romannumeral3}, the eRRHs coordinately transmit the cached and uncached requested files to the users.

The uncached files in the BBU are quantized and precoded and then delivered to the eRRHs via the fronthaul links. Let $\widetilde{\mathbf{x}}_{i}$ denote the precoded signal of the requested files that are not stored at eRRH $i$, which is given by
\begin{equation}\label{DelayAware05}
\widetilde{\mathbf{x}}_{i}=\sum\limits_{g\in \mathcal{G}}\overline{c}_{f_{g},i}\mathbf{v}_{g,i}\mathbf{s}_{f_{g}}
\end{equation}
where, $\mathbf{v}_{g,i}\in\mathds{C}^{N_{\mathrm{t}}\times 1}$ is the beamforming vector for the uncached requested file $f_{g}$ at eRRH $i$. Let $\widehat{\mathbf{x}}_{i}=\widetilde{\mathbf{x}}_{i}+\mathbf{z}_{i}$ be the quantized version of precoded signal $\widetilde{\mathrm{x}}_{i}$ at the BBU, where $\mathbf{z}_{i}\in\mathds{C}^{N_{\mathrm{t}}\times 1}$ is the quantization noise independent of $\widetilde{\mathbf{x}}_{i}$ with distribution $\mathbf{z}_{i}\sim\mathcal{CN}\left(\mathbf{0},\mathbf{\Omega}_{i}\right)$. We assume that the quantization noise $\mathbf{z}_{i}$ is independent across the eRRHs, i.e., the signals intended for different eRRHs are quantized independently~\cite{TVTKang2016}. Let $\mathbf{\Omega}$ be the covariance matrix of quantization noise, i.e.,  $\mathbf{\Omega}=\mathrm{Diag}\left(\mathbf{\Omega}_{1},\cdots,\mathbf{\Omega}_{K_{\mathrm{R}}}\right)$.

The signal $\mathbf{x}_{i}$ transmitted by eRRH $i$ is a superposition of two signals, where one is the locally precoded signal of the cached requested files and the other is the precoded and quantized signal of the uncached requested files stored at the BBU, which is delivered to eRRH $i$ via the error-free fronthaul link. Therefore, we have
\begin{equation}\label{DelayAware04}
\mathbf{x}_{i}=\sum\limits_{g\in \mathcal{G}}c_{f_{g},i}\mathbf{u}_{g,i}\mathbf{s}_{f_{g}}+\widehat{\mathbf{x}}_{i}
=\sum\limits_{g\in \mathcal{G}}c_{f_{g},i}\mathbf{u}_{g,i}\mathbf{s}_{f_{g}}+\sum\limits_{g\in \mathcal{G}}\overline{c}_{f_{g},i}\mathbf{v}_{g,i}\mathbf{s}_{f_{g}}+\mathrm{z}_{i}
\end{equation}
where $\mathbf{u}_{g,i}\in\mathds{C}^{N_{\mathrm{t}}\times 1}$ denotes the beamforming vectors for the cached requested file $f_{g}$ at eRRH $i$. The received baseband signal at user $k$ is expressed as
\begin{equation}\label{DelayAware07}
y_{k,2}=\sum\limits_{g\in\mathcal{G}}
\mathbf{h}_{k}^{H}\overline{\mathbf{w}}_{g}s_{f_{g}}+\mathbf{h}_{k}^{H}\mathbf{z}+n_{k}
\end{equation}
where $\overline{\mathbf{w}}_{g}=\mathbf{u}_{g}+\mathbf{v}_{g}$, $\mathbf{u}_{g}=\left[c_{f_{g},1}\mathbf{u}_{g,1}^{H},\cdots,c_{f_{g},K_{\mathrm{R}}}\mathbf{u}_{g,K_{\mathrm{R}}}^{H}
\right]^{H}$, $\mathbf{v}_{g}=\left[\overline{c}_{f_{g},1}\mathbf{v}_{g,1}^{H},
\cdots,\overline{c}_{f_{g},K_{\mathrm{R}}}\mathbf{v}_{g,K_{\mathrm{R}}}^{H}\right]^{H}$, and  $\mathbf{z}=\left[\left(\mathbf{z}_{1}\right)^{H},\cdots,\left(\mathbf{z}_{K_{\mathrm{R}}}
\right)^{H}\right]^{H}$. It is easy to see that $\overline{\mathbf{w}}_{g,i}=c_{f_{g},i}\mathbf{u}_{g,i}+\overline{c}_{f_{g},i}\mathbf{v}_{g,i}$. Furthermore, only one of $\overline{\mathbf{w}}_{g,i}=\mathbf{u}_{g,i}$ and $\overline{\mathbf{w}}_{g,i}=\mathbf{v}_{g,i}$ holds. Thus, the SINR at user $k$ is given by
\begin{equation}\label{DelayAware08}
\gamma_{k,2}=\frac{\left|\mathbf{h}_{k}^{H}\overline{\mathbf{w}}_{g_{k}}\right|^{2}}
{\sum\limits_{g\in\mathcal{G}\setminus\left\{g_{k}\right\}}
\left|\mathbf{h}_{k}^{H}\overline{\mathbf{w}}_{g}\right|^{2}+\mathbf{h}_{k}^{H}\mathbf{\Omega}\mathbf{h}_{k}+\sigma_{k}^{2}}.
\end{equation}
The corresponding achievable rate on unit bandwidth in one second of user $k$ is calculated as $R_{k,2}=\ln\left(1+\gamma_{k,2}\right)$ in unit of nats/Hz/s.

In general, fetching a requested file from the BBU via the fronthaul link incurs a certain delay since the propagation of fronthaul links and signal processing at the BBU. Such a delay is mainly determined by the worst transfer of the fronthaul links. We define the worst delay $\tau$, in unit of second, as follows\footnote{If eRRH $i\in\mathcal{K}_{\mathrm{R}}$ has cached all requested files at its local cache, i.e., $\sum\limits_{g\in\mathcal{G}}\overline{c}_{f_{g},i}=0$, the value of $\ln\left(\left|\mathbf{A}_{i}\right|\right)-\ln\left(\left|\mathbf{\Omega}_{i}\right|\right)$ in the denominator of the second item in~\eqref{DelayAware11} is set to be a very large constant value.}
\begin{equation}\label{DelayAware11}
\tau=\tau_{0}+\frac{S}{\min\limits_{i\in\mathcal{K}_{\mathrm{R}}}\left(\ln\left(\left|\mathbf{A}_{i}
\right|\right)-\ln\left(\left|\mathbf{\Omega}_{i}\right|\right)\right)}
\end{equation}
where $\tau_{0}$ denotes the constant delay for constant route time and signal processing at the BBU and $\mathbf{A}_{i}=\sum\limits_{g\in\mathcal{G}}\overline{c}_{f_{g},i}\mathbf{v}_{g,i}
\mathbf{v}_{g,i}^{\mathrm{H}}+\mathbf{\Omega}_{i}$. The second term in~\eqref{DelayAware11} accounts for the worst transfer delay of the propagation of fronthaul links. Consequently, the delivery latency minimization problem is formulated as follows
\begin{subequations}\label{DelayAware10}
\begin{align}
&\min\max_{g\in\mathcal{G}}\frac{S}
{\mathrm{r}_{g,2}}+\tau\label{DelayAware10a}\\
\mathrm{s.t.}~&\mathrm{r}_{{g},2}\leq\mathrm{R}_{k,2}, \forall k\in\mathcal{G}_{g}, \forall g\in\mathcal{G} \label{DelayAware10b}\\
&\sum\limits_{g\in\mathcal{G}}\left\|\overline{\mathbf{w}}_{g,i}\right\|^{2}+
\mathrm{Tr}\left(\mathbf{\Omega}_{i}\right)\leq P_{i}, \forall i\in\mathcal{K}_{\mathrm{R}}\label{DelayAware10c}\\
&g_{i}\left(\mathcal{V},\mathcal{O}\right)\leqslant C_{i}, \mathbf{\Omega}_{i}\succeq\mathbf{0}, \forall i\in\mathcal{K}_{\mathrm{R}}\label{DelayAware10d}
\end{align}
\end{subequations}
where the optimization variables are $\mathbf{u}_{g}$, $\mathbf{v}_{g}$, $\mathbf{\Omega}$, and $\mathrm{r}_{g,2}$, $\forall g\in\mathcal{G}$, $r_{g,2}$ in unit of nats/Hz/s denotes the delivery rate of group $g$, and $g_{i}\left(\mathcal{V},\mathcal{O}\right)$ denotes the rate on the fronthaul link of eRRH $i$ and is given by
\begin{equation}\label{DelayAware06}
g_{i}\left(\mathcal{V},\mathcal{O}\right)\triangleq I\left(\widetilde{\mathbf{x}}_{i}; \widehat{\mathbf{x}}_{i}\right)
=\ln\left(\left|\mathbf{A}_{i}\right|\right)-\ln\left(\left|\mathbf{\Omega}_{i}\right|\right)
\end{equation}
where $\mathcal{V}\triangleq\left\{\mathbf{v}_{g,i}\right\}_{g\in \mathcal{G},i\in\mathcal{K}_{\mathrm{R}}}$ and $\mathcal{O}\triangleq\left\{\mathbf{\Omega}_{i}\right\}_{i\in\mathcal{K}_{\mathrm{R}}}$. Constraints~\eqref{DelayAware10b} means that the delivery rate of file requested by the users in group $\mathcal{G}_{g}$ is no larger than the achievable rate of user $k$ that belongs to group $\mathcal{G}_{g}$. Constraints~\eqref{DelayAware10c} is the power constraint per eRRH. Constraint~\eqref{DelayAware10d} is the fronthaul capacity constraint ensuring signal $\widehat{\mathbf{x}}_{i}$ can be reliably recovered by eRRH $i$~\cite[Ch.~3]{BookGamal2011}. Note that when no requested files are cached at the eRRHs, i.e., $c_{f,i}=0$, $\forall i\in\mathcal{K}_{\mathrm{R}}$, $\forall f\in\mathcal{F}_{\mathrm{req}}$, the delivery latency minimization problem can still be formulated by~\eqref{DelayAware10}. When all requested files are stored at the eRRHs, i.e., $\sum\limits_{g\in\mathcal{G}}c_{f_{g},i}=\left|\mathcal{F}_{\mathrm{req}}\right|$, $\forall i\in\mathcal{K}_{\mathrm{R}}$, the PCBT scheme reduces to the FCBT scheme, i.e., problem~\eqref{DelayAware10} is equivalent to problem~\eqref{DelayAware09}.

\subsection*{C. Partial Caching Pipelined Transmission Scheme}

\begin{figure}[t]
\renewcommand{\captionfont}{\footnotesize}
\renewcommand*\captionlabeldelim{.}
	\centering
	\captionstyle{flushleft}
	\onelinecaptionstrue
	\includegraphics[width=1\columnwidth,keepaspectratio]{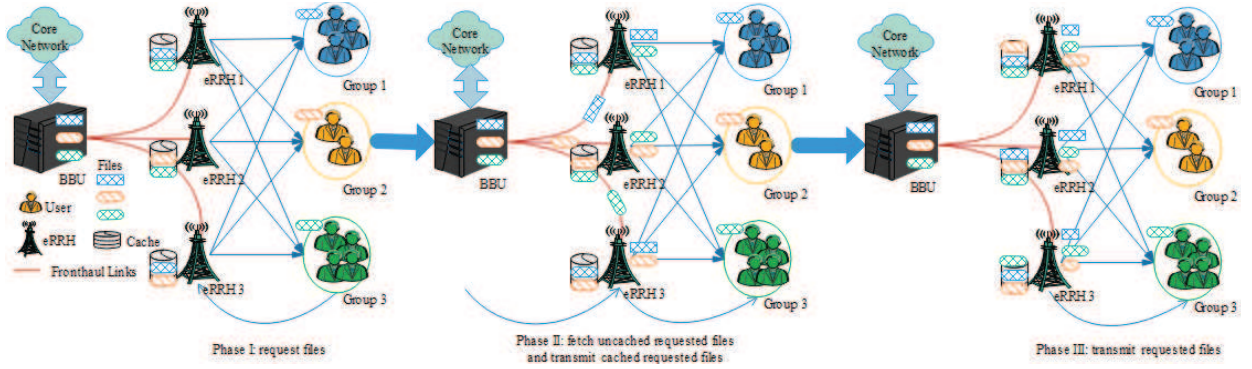}\\
	\caption{Flowchart of partial caching pipelined transmission scheme.}
	\label{TwoPhaseFlowChart}
\end{figure}

In the PCBT scheme, the eRRHs have to wait for the arrival of the uncached requested files before transmitting all requested files to the users, that is, waiting for delay $\tau$. In practice, an eRRH is able to receive data from its fronthaul link while sending wireless signals. Thus, we design a novel partial caching pipelined transmission (PCPT) scheme that also contains three phases, as shown in Fig.~\ref{TwoPhaseFlowChart}. Specifically, in phase \uppercase\expandafter{\romannumeral1}, the users first send the requirement of files to the eRRHs. After receiving the requirements of the users, in phase \uppercase\expandafter{\romannumeral2}, according to the caching status of the requested files, the eRRHs transmit the cached requested files to the users while fetching the uncached requested files from the BBU. In phase \uppercase\expandafter{\romannumeral3}, after the arrival of the uncached requested files, the eRRHs transmit the remaining cached requested files and uncached requested files to the users. Different from the PCBT scheme, the eRRHs do not have to wait for the uncached requested files to arrive before sending the cached requested files.

The duration of phase \uppercase\expandafter{\romannumeral2} is delay $\tau$ given by~\eqref{DelayAware11}, i.e., the time of fetching the uncached requested files from the BBU and the signal processing at the BBU, etc. In phase \uppercase\expandafter{\romannumeral2}, the eRRHs cooperatively transmit the cached requested files to the users. Thus, the received signal at each user is expressed as~\eqref{DelayAware02}. In phase \uppercase\expandafter{\romannumeral3}, after the quantized precoded signals of the uncached requested files arrives at all eRRHs, the remaining cached requested files and uncached requested files are simultaneously transmitted to all users as in the PCBT scheme. Hence, the received signal at each user is expressed  as~\eqref{DelayAware07}. Consequently, the delivery latency minimization problem is formulated as\footnote{The existing work in~\cite{TCOMHe2018} maximizes the minimum delivery rate with fixed delay $\tau$ incurred by the propagation of fronthaul links and signal processing at the BBU. However, this work aims to minimize the delivery latency and optimize the delay $\tau$. As a consequence, the problem considered in this paper is more comprehensive as compared with that in~\cite{TCOMHe2018}.}
\begin{subequations}\label{DelayAware14}
\begin{align}
&\min\max_{g\in\mathcal{G}}\frac{S-\tau\mathrm{r}_{g,1}}
{\mathrm{r}_{g,2}}+\tau\label{DelayAware14a}\\
\mathrm{s.t.}~&\eqref{DelayAware09b},\eqref{DelayAware09c},\eqref{DelayAware10b},
\eqref{DelayAware10c},\eqref{DelayAware10d}\label{DelayAware14b}\\
&\tau\mathrm{r}_{g,1}\leq S, \forall g\in\mathcal{G} \label{DelayAware14c}
\end{align}
\end{subequations}
where the optimization variables are $\mathbf{w}_{g}$, $\mathbf{u}_{g}$, $\mathbf{v}_{g}$, $\mathbf{\Omega}$, and $\mathrm{r}_{g,p}$, $\forall g\in\mathcal{G}$, $\forall p\in\mathcal{P}=\left\{1,2\right\}$. In~\eqref{DelayAware14a}, $S-\tau\mathrm{r}_{g,1}$ denotes the remaining cached requested files after the transmission of phase \uppercase\expandafter{\romannumeral2}. Constraints~\eqref{DelayAware14c} imposes that the amount of data transmission of each file be limited by file size $S$. Note that in problem~\eqref{DelayAware14}, when all requested files are locally cached by the eRRHs, i.e., $\sum\limits_{g\in\mathcal{G}}c_{f_{g},i}
=\left|\mathcal{F}_{\mathrm{req}}\right|$, $\forall i\in\mathcal{K}_{\mathrm{R}}$, the value of $\tau$ is zero and problem~\eqref{DelayAware14} is equivalent to problem~\eqref{DelayAware09}. When no requested files are stored at the cache of the network edge, i.e., $\sum\limits_{g\in\mathcal{G}}\sum\limits_{i\in\mathcal{K}_{\mathrm{R}}}c_{f_{g},i}
=0$, problem~\eqref{DelayAware14} reduces to problem~\eqref{DelayAware10}. When requested file $f_{g}$ is not cached at the eRRHs, i.e., $c_{f_{g},i}=0$, $\forall i\in\mathcal{K}_{\mathrm{R}}$, constraints $\mathrm{r}_{g,1}\leq\mathrm{R}_{k,1}$, $\forall k\in\mathcal{G}_{g}$, and $\tau\mathrm{r}_{g,1}\leq S$ corresponding to group $g$ in~\eqref{DelayAware09b} and~\eqref{DelayAware14c} are redundant. 
\section*{\sc \uppercase\expandafter{\romannumeral4}. Design of Optimization Algorithms}
In this section, we focus on designing an efficient optimization algorithm to solve the corresponding optimization problem of each transmission scheme.
The common characteristics of these problems are the non-convex group rate constraints~\eqref{DelayAware09b} and~\eqref{DelayAware10b}, due to the existence of non-convex achievable rate in the right side of constraints~\eqref{DelayAware09b} and~\eqref{DelayAware10b}. Problems~\eqref{DelayAware10} and~\eqref{DelayAware14} are even more challenging due to the non-convex fractional objective function and fronthaul capacity constraints. Therefore, problems~\eqref{DelayAware09},~\eqref{DelayAware10} and~\eqref{DelayAware14} are non-convex and it is difficult to obtain the global optimum. In what follows, we design heuristic optimization algorithms to solve the problems locally via the penalty dual decomposition (PDD) method and the successive convex approximation (SCA) methods~\cite{FoundBoyd2011,JSACTsinos2017,JSTSPShi2018,TSPSun2018}.

\subsection*{A. Solving Problem~\eqref{DelayAware09} FCBT Scheme}
In this subsection, we focus on solving problem~\eqref{DelayAware09}. The main barriers of solving problem~\eqref{DelayAware09} are the non-convexity of the objective function~\eqref{DelayAware09a} and the group rate constraints~\eqref{DelayAware09b}. First, we need to convert the non-convex forms into convex ones. Introducing auxiliary variables $\overline{\gamma}_{k,1}$, $\chi_{k,1}$, $\forall k\in\mathcal{G}_{g}$, $\forall g\in\mathcal{G}$, problem~\eqref{DelayAware09} can be equivalently reformulated into the following
\begin{subequations}\label{DelayAware15}
\begin{align}
&\min\max_{g\in\mathcal{G}}\frac{S}
{\mathrm{r}_{g,1}}\label{DelayAware15a}\\
\mathrm{s.t.}~&\mathrm{r}_{g,1}\leq\ln\left(1+\overline{\gamma}_{k,1}\right), \forall k\in\mathcal{G}_{g}, \forall g\in\mathcal{G}\label{DelayAware15b}\\
&\overline{\gamma}_{k,1}\leq\frac{\left|\mathbf{h}_{k}^{H}\mathbf{w}_{g_{k}}\right|^{2}}{\chi_{k,1}}, \forall k\in\mathcal{K}_{\mathrm{U}}\label{DelayAware15c}\\
&\sum\limits_{g\in\mathcal{G}\setminus\left\{g_{k}\right\}}
\left|\mathbf{h}_{k}^{H}\mathbf{w}_{g}\right|^{2}+\sigma_{k}^{2}\leq\chi_{k,1}, \forall k\in\mathcal{K}_{\mathrm{U}}\label{DelayAware15d}\\
&\sum\limits_{g\in\mathcal{G}}\left\|c_{f_{g},i}\mathbf{w}_{g,i}\right\|^{2}\leq P_{i}, \forall i\in\mathcal{K}_{\mathrm{R}}\label{DelayAware15e}
\end{align}
\end{subequations}
where the optimization variables are $\mathbf{w}_{g}$, $\mathrm{r}_{g,1}$, $\forall g\in\mathcal{G}$, $\overline{\gamma}_{k,1}$, and $\chi_{k,1}$, $\forall k\in\mathcal{K}_{\mathrm{U}}$. At the optimal point of problem~\eqref{DelayAware15}, the inequality constraints~\eqref{DelayAware15c} and~\eqref{DelayAware15d} are activated. In~\eqref{DelayAware15c}, $\overline{\gamma}_{k,1}$ and $\frac{\left|\mathbf{h}_{k}^{H}\mathbf{w}_{g_{k}}\right|^{2}}{\chi_{k,1}}$, $\forall k\in\mathcal{K}_{\mathrm{U}}$, are convex, respectively. But, constraints~\eqref{DelayAware15c} is still non-convex. To deal with the non-convex constraints, we invoke a result of~\cite{OperalMarks1977,JBeck2010,SPLTran2012} which shows that if we replace $\frac{\left|\mathbf{h}_{k}^{H}\mathbf{w}_{g_{k}}\right|^{2}}{\chi_{k,1}}$ by its convex low bound and iteratively solve the resulting problem by judiciously updating the variables until
convergence, we can obtain a Karush-Kuhn-Tucker (KKT) point of problem~\eqref{DelayAware15}. To this end,  we approximate problem~\eqref{DelayAware15} as follows
\begin{subequations}\label{DelayAware16}
\begin{align}
&\min\max_{g\in\mathcal{G}}\frac{S}
{\mathrm{r}_{g,1}}\label{DelayAware16a}\\
\mathrm{s.t.}~&\eqref{DelayAware15b}, \eqref{DelayAware15d}, \eqref{DelayAware15e} \label{DelayAware16b}\\
&\overline{\gamma}_{k,1}\leq\overline{\varphi}^{\left(t\right)}
\left(\mathbf{w}_{g_{k}},\chi_{k,1}\right)
, \forall k\in\mathcal{K}_{\mathrm{U}}\label{DelayAware16c}
\end{align}
\end{subequations}
where the optimization variables are $\mathbf{w}_{g}$, $\mathrm{r}_{g,1}$, $\forall g\in\mathcal{G}$, $\overline{\gamma}_{k,1}$, and $\chi_{k,1}$, $\forall k\in\mathcal{K}_{\mathrm{U}}$. In~\eqref{DelayAware16c}, $\overline{\varphi}^{\left(t\right)}\left(\mathbf{w},\chi\right)$ is a convex low boundary of function $\frac{\left|\mathbf{h}_{k}^{H}\mathbf{w}_{g_{k}}\right|^{2}}{\chi_{k,1}}$ and is defined as
\begin{equation}\label{DelayAware17}
\overline{\varphi}^{\left(t\right)}\left(\mathbf{w},\chi\right)
\triangleq\frac{2\Re\left(\left(\mathbf{w}^{\left(t\right)}
\right)^{\mathrm{H}}\mathbf{h}_{k}\mathbf{h}_{k}^\mathrm{H}	\mathbf{w}\right)}{\chi^{\left(t\right)}}
-\left(\frac{\left|\mathbf{h}_{k}^{H}\mathbf{w}^{\left(t\right)}\right|}
{\chi^{\left(t\right)}}\right)^{2}\chi
\end{equation}
where $t$ denotes the index of iteration, $\mathbf{w}^{\left(t\right)}$ and $\chi^{\left(t\right)}$ represent the values of variables $\mathbf{w}$ and $\chi$ obtained at the $t$-th iteration, respectively.

Next, we pay our attention to objective function~\eqref{DelayAware16a}. By introducing auxiliary variable $\eta$, we can transform problem~\eqref{DelayAware16} equivalently into the following convex form
\begin{subequations}\label{DelayAware18}
\begin{align}
&\min~\eta\label{DelayAware18a}\\
\mathrm{s.t.}~&\eqref{DelayAware15b}, \eqref{DelayAware15d}, \eqref{DelayAware15e}, \eqref{DelayAware16c} \label{DelayAware18b}\\
&\ln\left(S\right)-\ln\left(\eta\right)-\ln\left(\mathrm{r}_{g,1}\right)\leq 0, \forall g\in\mathcal{G} \label{DelayAware18c}
\end{align}
\end{subequations}
where the optimization variables are $\eta$, $\mathbf{w}_{g}$, $\mathrm{r}_{g,1}$, $\forall g\in\mathcal{G}$, $\overline{\gamma}_{k,1}$ and $\chi_{k,1}$, $\forall k\in\mathcal{K}_{\mathrm{U}}$. Note that in constraint~\eqref{DelayAware18c}, we exploit the positive nature of $\eta$ and $r_{g,1}$, i.e., $\eta>0$ and $r_{g,1}>0$. This is because if $r_{g,1}=0$, the delivery latency is infinite, i.e., problem~\eqref{DelayAware18} becomes meaningless. At the $\left(t+1\right)$-th iteration, problem~\eqref{DelayAware18} is convex and can be easily solved with a classical optimization solver, such as CVX~\cite{MathPotra2000,AvialGrant2015}. The detailed steps of solving problem~\eqref{DelayAware18} are summarized in Algorithm~\ref{DelayAwareAlg01} that converges to a KKT solution of problem~\eqref{DelayAware10}, please see Appendix A for the detailed proof.

\begin{algorithm}[t]
\caption{Solution of problem~\eqref{DelayAware18}}\label{DelayAwareAlg01}
\begin{algorithmic}[1]
\STATE Set $t=0$ and $\eta^{\left(t\right)}$ to a non-zero value. Initializing $\mathbf{w}_{g}^{\left(t\right)}$ to be a non-zero beamforming vector, $\forall g\in\mathcal{G}$, such that constraint \eqref{DelayAware14c} is satisfied; \label{DelayAwareAlg0101}
\STATE Compute $\chi_{k,1}^{\left(t\right)}$ as follows\label{DelayAwareAlg0102}
\begin{equation}\label{DelayAware19}
\chi_{k,1}^{\left(t\right)}=\sum\limits_{g\in\mathcal{G}\setminus\left\{f_{g_{k}}\right\}}
\left|\mathbf{h}_{k}^{H}\mathbf{w}_{g}^{\left(t\right)}\right|^{2}+\sigma_{k}^{2}, \forall k\in\mathcal{K}_{\mathrm{U}};
\end{equation}
\STATE Solve problem~\eqref{DelayAware18} to obtain $\eta^{\left(t+1\right)}$, $\mathbf{w}_{g}^{\left(t+1\right)}$, $\mathrm{r}_{g,1}^{\left(t+1\right)}$, $\forall g\in\mathcal{G}$, $\overline{\gamma}_{k,1}^{\left(t+1\right)}$ and $\chi_{k,1}^{\left(t+1\right)}$, $\forall k\in\mathcal{K}_{\mathrm{U}}$;\label{DelayAwareAlg0103}
\STATE If $\left|\frac{\eta^{\left(t+1\right)}-\eta^{\left(t\right)}}{\eta^{\left(t\right)}}\right|\leqslant\varepsilon$, stop iteration. Otherwise, set $t\leftarrow t+1$ and go to Step~\ref{DelayAwareAlg0102}.\label{DelayAwareAlg0104}
\end{algorithmic}
\end{algorithm}

\subsection*{B. Solving Problem~\eqref{DelayAware10} for PCBT Scheme}
In this subsection, we focus on investigating the solution of problem~\eqref{DelayAware10}, and propose an effecient optimization method to solve it. Compared to problem~\eqref{DelayAware09}, solving problem~\eqref{DelayAware10} becomes more challenging, because there are additional non-convex fractional item in the objective~\eqref{DelayAware10a} and non-convex fronthaul capacity constraints~\eqref{DelayAware10d}. To overcome these difficulties, we need to leverage some new mathematical methods to transform non-convex problem~\eqref{DelayAware10} into convex one. For simplicity, define $\mathbf{H}_{k}=\mathbf{h}_{k}\mathbf{h}_{k}^{H}$, $\forall k\in\mathcal{K}_{\mathrm{U}}$, and $\overline{\mathbf{W}}_{g}=\overline{\mathbf{w}}_{g}\overline{\mathbf{w}}_{g}^{H}$, $\forall g\in\mathcal{G}$. Note that $\overline{\mathbf{W}}_{g}=\overline{\mathbf{w}}_{g}\overline{\mathbf{w}}_{g}^{H}$ if and only if $\overline{\mathbf{W}}_{g}=\overline{\mathbf{w}}_{g}\overline{\mathbf{w}}_{g}^{H}\succeq\mathbf{0}$ and $\mathrm{rank}\left(\overline{\mathbf{W}}_{g}\right)=1$. Dropping the rank one constraint of $\overline{\mathbf{W}}_{g}$, $\forall g\in\mathcal{G}$, problem~\eqref{DelayAware10} can be rewritten as
\begin{subequations}\label{DelayAware20}
\begin{align}
&\min\max_{g\in\mathcal{G}}\frac{S}
{\mathrm{r}_{g,2}}+\tau\label{DelayAware20a}\\
\mathrm{s.t.}~&\mathrm{r}_{g,2}-\ln\left(\mu_{k,2}\right)
+\ln\left(\chi_{k,2}\right)\leq 0, \forall k\in\mathcal{G}_{g}, \forall g\in\mathcal{G}\label{DelayAware20b}\\
&\sum\limits_{g\in\mathcal{G}}\mathrm{Tr}\left(\mathbf{P}_{g,i}^{T}\left(1\right)
\overline{\mathbf{W}}_{g}\mathbf{P}_{g,i}\left(1\right)\right)
+\mathrm{Tr}\left(\mathbf{\Omega}_{i}\right)\leq P_{i}, \forall i\in\mathcal{K}_{\mathrm{R}}\label{DelayAware20c}\\
&\overline{\mathbf{W}}_{g}\succeq \mathbf{0},\forall g\in\mathcal{G}, \mathbf{\Omega}_{i}\succeq \mathbf{0},\forall i\in\mathcal{K}_{\mathrm{R}}\label{DelayAware20d}\\
&g_{i}\left(\mathcal{V},\mathcal{O}\right)\leqslant C_{i}, \forall i\in\mathcal{K}_{\mathrm{R}}\label{DelayAware20e}
\end{align}
\end{subequations}
where the optimization variables are $\overline{\mathbf{W}}_{g}$, $\mathbf{\Omega}$, $\mathrm{r}_{g,2}$, $\forall g\in\mathcal{G}$. In~\eqref{DelayAware20b}, $\mu_{k,2}$ and $\chi_{k,2}$ are defined respectively by
\begin{subequations}\label{DelayAware21}
\begin{align}
\mu_{k,2}&=\sum\limits_{g\in\mathcal{G}}
\mathrm{Tr}\left(\mathbf{H}_{k}\overline{\mathbf{W}}_{g}\right)+\mathrm{Tr}\left(
\mathbf{\Omega}\mathbf{H}_{k}\right)+\sigma_{k}^{2}, \forall k\in\mathcal{G}_{g}, \forall g\in\mathcal{G}\\
\chi_{k,2}&=\sum\limits_{g'\in\mathcal{G}\setminus\left\{g\right\}}
\mathrm{Tr}\left(\mathbf{H}_{k}\overline{\mathbf{W}}_{g'}\right)+\mathrm{Tr}\left(
\mathbf{\Omega}\mathbf{H}_{k}\right)+\sigma_{k}^{2}, \forall k\in\mathcal{G}_{g}, \forall g\in\mathcal{G}.
\end{align}
\end{subequations}
In~\eqref{DelayAware20c}, permutation matrix function $\mathbf{P}_{g,i}\left(x\right)$ is defined as
\begin{equation}\label{DelayAware22}
\mathbf{P}_{g,i}\left(x\right)=\left[\mathbf{0}_{N_{\mathrm{t}}\times \left(i-1\right)N_{\mathbf{t}}},x\mathbf{I}_{N_{\mathrm{t}}\times N_{\mathrm{t}}},\mathbf{0}_{N_{\mathrm{t}}\times \left(K_{\mathrm{R}}N_{\mathbf{t}}-iN_{\mathbf{t}}\right)}\right]^{T}.
\end{equation}
Problem~\eqref{DelayAware20} is non-convex due to the non-convexity of objective function~\eqref{DelayAware20a} and constraints~\eqref{DelayAware20b} and~\eqref{DelayAware20e}. Consequently, it is difficult to obtain the global optimal solution of problem~\eqref{DelayAware20}. In what follows, we aim to relax the optimization conditions in order to provide reasonable design for practical implementation.

The first thing of addressing problem~\eqref{DelayAware20} is to transfer it into a solvable form by using some mathematical methods. By introducing auxiliary variables $\eta$ and $\theta$, problem~\eqref{DelayAware20} can be equivalently reformulated as
\begin{subequations}\label{DelayAwarer26}
\begin{align}
&\min~\eta+\theta\label{DelayAwarer26a}\\
\mathrm{s.t.}~&\eqref{DelayAware20b},\eqref{DelayAware20c}, \eqref{DelayAware20d},\eqref{DelayAware20e}, \label{DelayAwarer26b}\\
&S\leq \eta\mathrm{r}_{g,2}, \forall g\in\mathcal{G} \label{DelayAwarer26c}\\
&\theta=\frac{S}{\min\limits_{i\in\mathcal{K}_{\mathrm{R}}}\left(\ln\left(\left|\mathbf{A}_{i}
\right|\right)-\ln\left(\left|\mathbf{\Omega}_{i}\right|\right)\right)} \label{DelayAwarer26d}
\end{align}
\end{subequations}
where the optimization variables are $\eta$, $\theta$, $\overline{\mathbf{W}}_{g}$, $\mathbf{\Omega}$, $\mathrm{r}_{g,2}$, $\forall g\in\mathcal{G}$. Note that constant $\tau_{0}$ in objective function~\eqref{DelayAwarer26a} is omitted. Problem~\eqref{DelayAwarer26} can be convexified as problem~\eqref{DelayAwarer30} via some basic mathematical operation and using the PDD and SCA methods~\cite{FoundBoyd2011,JSACTsinos2017,JSTSPShi2018,TSPSun2018}, please see Appendix B for the details.
\begin{subequations}\label{DelayAwarer30}
\begin{align}
&\min~\eta+\theta+\frac{1}{2\rho}\sum\limits_{i\in\mathcal{K}_{\mathrm{R}}}
\mathds{1}\left(\sum\limits_{g\in\mathcal{G}}\overline{c}_{f_{g},i}\right)\left|\frac{S}{\theta}+
\phi\left(\mathbf{\Omega}_{i},\mathbf{\Omega}_{i}^{\left(t\right)}\right)-
\ln\left(\left|\mathbf{A}_{i}\right|\right)+\rho\lambda\right|^{2}\label{DelayAwarer30a}\\
\mathrm{s.t.}~&\mathrm{r}_{g,2}-\ln\left(\mu_{k,2}\right)+\phi\left(\chi_{k,2},\chi_{k,2}^{\left(t\right)}\right)
\leq 0, \forall k\in\mathcal{K}_{\mathrm{U}},\label{DelayAwarer30b}\\
&\sum\limits_{g\in\mathcal{G}}\mathrm{Tr}\left(\mathbf{P}_{g,i}^{T}\left(1\right)
\overline{\mathbf{W}}_{g}\mathbf{P}_{g,i}\left(1\right)\right)
+\mathrm{Tr}\left(\mathbf{\Omega}_{i}\right)\leq P_{i}, \forall i\in\mathcal{K}_{\mathrm{R}}\label{DelayAwarer30c}\\
&\overline{\mathbf{W}}_{g}\succeq \mathbf{0},\forall g\in\mathcal{G}, \mathbf{\Omega}_{i}\succeq \mathbf{0},\forall i\in\mathcal{K}_{\mathrm{R}}\label{DelayAwarer30d}\\
&\phi\left(\mathbf{A}_{i},\mathbf{A}_{i}^{\left(t\right)}\right)
-\ln\left(\left|\mathbf{\Omega}_{i}\right|\right)\leqslant C_{i}, \forall i\in\mathcal{K}_{\mathrm{R}},\label{DelayAwarer30e}\\
&\ln\left(S\right)-\ln\left(\eta\right)-\ln\left(\mathrm{r}_{g,2}\right)\leq 0, \forall g\in\mathcal{G}, \label{DelayAwarer30f}
\end{align}
\end{subequations}
where the optimization variables are $\eta$, $\theta$, $\overline{\mathbf{W}}_{g}$, $\mathbf{\Omega}$, $\mathrm{r}_{g,2}$, $\forall g\in\mathcal{G}$. In problem~\eqref{DelayAwarer30}, $\lambda$ is the Lagrange multiplier and $\rho$ is a scalar penalty parameter. This penalty parameter improves the robustness compared to other optimization methods for constrained problems (e.g. dual ascent method) and in particular achieves convergence without the need of specific assumptions for the objective function, i.e. strict convexity and finiteness~\cite{FoundBoyd2011,JSACTsinos2017,JSTSPShi2018,TSPSun2018}\footnote{In constraint~\eqref{DelayAware26c}, we exploit the positive nature of $\eta$ and $r_{g,2}$, i.e., $\eta>0$ and $r_{g,2}>0$. This is because that if $r_{g,2}=0$, the delivery latency is infinite, i.e., problem~\eqref{DelayAware20} becomes meaningless.}.

When Lagrange multiplier $\lambda$ and penalty parameter $\rho$ are fixed, problem~\eqref{DelayAwarer30} is convex and can be easily solved by a classical optimization solver, such as the CVX~\cite{MathPotra2000,AvialGrant2015}. Based on this observation, in the sequel, we adopt an alternative optimization method to address problem~\eqref{DelayAwarer30}. In particular, we first solve problem~\eqref{DelayAwarer30} with fixed $\lambda$ and $\rho$, and then update Lagrange multiplier $\lambda$ and penalty parameter $\rho$ according to the constraint violation condition~\cite{JSTSPShi2018}. A step-by-step description for solving problem~\eqref{DelayAwarer30} is given in Algorithm~\ref{DelayAwareAlg02}, where $l$ and $t$ denote the number of iterations, respectively. $\epsilon$ and $\varsigma^{\left(l\right)}$ are a stopping threshold and an approximation stopping threshold, respectively. $\omega$ is a control parameter. $\zeta^{\left(t\right)}$ denotes the objective value of problem~\eqref{DelayAwarer30} at the $t$-th iteration.
\begin{algorithm}[t]
\caption{Solution of problem~\eqref{DelayAwarer30}}\label{DelayAwareAlg02}
\begin{algorithmic}[1]
\STATE Set $\zeta^{\left(0\right)}$ to be a non-zero value and initialize non-zero beamforming matrix $\overline{\mathbf{W}}_{g}^{\left(0\right)}$, $\forall g\in\mathcal{G}$, $\mathbf{\Omega}_{i}^{\left(0\right)}$, $\forall i\in\mathcal{K}_{\mathrm{R}}$, such that constraints~\eqref{DelayAware20c},~\eqref{DelayAware20d} and~\eqref{DelayAware20e} are satisfied; \label{DelayAwareAlg0201}
\STATE Let $l=0$, initialize $\lambda^{\left(l\right)}$ and $\rho^{\left(l\right)}$ to be a non-zero value;\label{DelayAwareAlg0202}
\STATE Let $t=0$, compute $\chi_{k,2}^{\left(t\right)}$ as follows:\label{DelayAwareAlg0203}
\begin{equation}\label{DelayAware31}
\chi_{k,2}^{\left(t\right)}=\sum\limits_{g\in\mathcal{G}\setminus\left\{g_{k}\right\}}
\mathrm{Tr}\left(\mathbf{H}_{k}\overline{\mathbf{W}}_{g}^{\left(t\right)}\right)+\mathrm{Tr}\left(
\mathbf{\Omega}^{\left(t\right)}\mathbf{H}_{k}\right)+\sigma_{k}^{2}, \forall k\in\mathcal{K}_{\mathrm{U}}.
\end{equation}
\STATE Let $t\leftarrow t+1$. Solve problem~\eqref{DelayAwarer30} to obtain $\zeta^{\left(t\right)}$, $\eta^{\left(t\right)}$, $\theta^{\left(t\right)}$, $\overline{\mathbf{W}}_{g}^{\left(t\right)}$, $\mathrm{r}_{g,2}^{\left(t\right)}$, $\forall g\in\mathcal{G}$, $\forall k\in\mathcal{K}_{\mathrm{U}}$, $\mathbf{\Omega}_{i}^{\left(t\right)}$, $\forall i\in\mathcal{K}_{\mathrm{R}}$;\label{DelayAwareAlg0204}
\STATE If $\left|\frac{\zeta^{\left(t\right)}-\zeta^{\left(t-1\right)}}{\zeta^{\left(t-1\right)}}\right|\leq\epsilon^{\left(l\right)}$, go to Step~\ref{DelayAwareAlg0206}. Otherwise, compute $\chi_{k,2}^{\left(t\right)}$, $\forall k\in\mathcal{K}_{\mathrm{U}}$, and go to Step~\ref{DelayAwareAlg0204};\label{DelayAwareAlg0205}
\STATE If $\left|\frac{S}{\theta^{\left(t\right)}}
-\min\limits_{i\in\mathcal{K}_{\mathrm{R}}}\left(\ln\left(\left|\mathbf{A}_{i}^{\left(t\right)}
\right|\right)-\ln\left(\left|\mathbf{\Omega}_{i}^{\left(t\right)}\right|\right)\right)\right|\leq\varepsilon$, stop iteration. Otherwise, go to Step~\ref{DelayAwareAlg0207};\label{DelayAwareAlg0206}
\STATE If $\left|\frac{S}{\theta^{\left(t\right)}}
-\min\limits_{i\in\mathcal{K}_{\mathrm{R}}}\left(\ln\left(\left|\mathbf{A}_{i}^{\left(t\right)}
\right|\right)-\ln\left(\left|\mathbf{\Omega}_{i}^{\left(t\right)}
\right|\right)\right)\right|\leq\varsigma^{\left(l\right)}$, update $\lambda$ and $\rho$ as follows\label{DelayAwareAlg0207}
\begin{subequations}\label{DelayAware32}
\begin{align}
\lambda^{\left(l+1\right)}&=\lambda^{\left(l\right)}+\frac{1}{\rho}\left(\frac{S}{\theta^{\left(t\right)}}
-\min\limits_{i\in\mathcal{K}_{\mathrm{R}}}\left(\ln\left(\left|\mathbf{A}_{i}^{\left(t\right)}
\right|\right)-\ln\left(\left|\mathbf{\Omega}_{i}^{\left(t\right)}\right|\right)\right)\right)\\
\rho^{\left(l+1\right)}&=\rho^{\left(l\right)}.
\end{align}
\end{subequations}
Otherwise, update $\lambda$ and $\rho$ as follows
\begin{subequations}\label{DelayAware33}
\begin{align}
\lambda^{\left(l+1\right)}&=\lambda^{\left(l\right)}\\
\rho^{\left(l+1\right)}&=\omega\rho^{\left(l\right)};
\end{align}
\end{subequations}
\STATE Let $l\leftarrow l+1$, $\varsigma^{\left(l+1\right)}=\omega\left|\frac{S}{\theta^{\left(t\right)}}
-\min\limits_{i\in\mathcal{K}_{\mathrm{R}}}\left(\ln\left(\left|\mathbf{A}_{i}^{\left(t\right)}
\right|\right)-\ln\left(\left|\mathbf{\Omega}_{i}^{\left(t\right)}
\right|\right)\right)\right|$, and go to Step~\ref{DelayAwareAlg0203}.\label{DelayAwareAlg0208}
\end{algorithmic}
\end{algorithm}
According to~\cite[Corollary 3.1]{JSTSPShi2018}, Algorithm~\ref{DelayAwareAlg02} guarantees convergence to a KKT solution of problem~\eqref{DelayAwarer26}. In Algorithm~\ref{DelayAwareAlg02}, Step~\ref{DelayAwareAlg0204} solves a convex problem, which can be efficiently implemented by the primal-dual interior point method with approximate complexity of $\mathit{O}\left(\left(G\left(2N_{\mathrm{t}}^{2}+1\right)+2\right)^{3.5}\right)$~\cite{MathPotra2000}. The overall computational complexity of Algorithm~\ref{DelayAwareAlg02} is $\mathit{O}\left(\upsilon_{2}\left(G\left(N_{\mathrm{t}}K_{\mathrm{R}}+4\right)\right)^{3.5}\right)$, where $\upsilon_{2}$ denotes the number of the operations of Step~\ref{DelayAwareAlg0204}. Due the influence of the rank relaxation, an optimal solution of problem~\eqref{DelayAwarer30} is not necessary an optimal solution of problem~\eqref{DelayAware20}. Therefore, we need to adopt a specific method that can be found in Appendix C to obtain the solution of problem~\eqref{DelayAware20} from the solution of problem~\eqref{DelayAwarer30}. The initialization of Algorithm~\ref{DelayAwareAlg02} is finished using the method proposed in Appendix D.

\subsection*{C. Solving Problem~\eqref{DelayAware14} for PCPT Scheme}

In this subsection, we focus on the optimization of problem~\eqref{DelayAware14} for the PCPT scheme, under the assumption that partial requested files are cached at the local cache of the network edge and partial requested files need to be fetched from the BBU. Compared to problems~\eqref{DelayAware09} and~\eqref{DelayAware10}, solving problem~\eqref{DelayAware14} is more challenging as the pipelined transmission of requested files. Following the similar procedure used for problem~\eqref{DelayAware10}, problem~\eqref{DelayAware14} can be reformulated as
\begin{subequations}\label{DelayAware35}
\begin{align}
&\min\max_{g\in\mathcal{G}}\frac{S-\tau\mathrm{r}_{g,1}}
{\mathrm{r}_{g,2}}+\tau\label{DelayAware35a}\\
\mathrm{s.t.}~&\eqref{DelayAware20b}, \eqref{DelayAware20c}, \eqref{DelayAware20d}, \eqref{DelayAware20e}\label{DelayAware35b}\\
&\mathrm{r}_{g,1}-\ln\left(\mu_{k,1}\right)
+\ln\left(\chi_{k,1}\right)\leq 0, \forall k\in\mathcal{G}_{g}, \forall g\in\mathcal{G}\label{DelayAware35c}\\
&\sum\limits_{g\in\mathcal{G}}\mathrm{Tr}\left(\mathbf{P}_{g,i}^{T}\left(c_{f_{g},i}\right)
\mathbf{W}_{g}\mathbf{P}_{g,i}\left(c_{f_{g},i}\right)\right)\leq P_{i}, \forall i\in\mathcal{K}_{\mathrm{R}}\label{DelayAware35d}\\
&\tau\mathrm{r}_{g,1}\leq S, \forall g\in\mathcal{G} \label{DelayAware35e}
\end{align}
\end{subequations}
where the optimization variables are $\mathbf{W}_{g}$, $\overline{\mathbf{W}}_{g}$, $\mathbf{\Omega}$, $\mathrm{r}_{g,p}$, $\forall g\in\mathcal{G}$, $\forall p\in\mathcal{P}$. In~\eqref{DelayAware35c}, $\mu_{k,1}$ and $\chi_{k,1}$ are defined respectively as
\begin{subequations}\label{DelayAware36}
\begin{align}
\mu_{k,1}&=\sum\limits_{g\in\mathcal{G}}
\mathrm{Tr}\left(\mathbf{H}_{k}\mathbf{W}_{g}\right)+\sigma_{k}^{2},\forall k\in\mathcal{G}_{g}, \forall g\in\mathcal{G}\\
\chi_{k,1}&=\sum\limits_{g'\in\mathcal{G}\setminus\left\{g\right\}}
\mathrm{Tr}\left(\mathbf{H}_{k}\mathbf{W}_{g'}\right)+\sigma_{k}^{2},\forall k\in\mathcal{G}_{g}, \forall g\in\mathcal{G}
\end{align}
\end{subequations}
where $\mathbf{W}_{g}=\mathbf{w}_{g}\mathbf{w}_{g}^{H}$, $\forall g\in\mathcal{G}$. $\mathbf{W}_{g}=\mathbf{w}_{g}\mathbf{w}_{g}^{H}$ if and only if $\mathbf{W}_{g}\succeq\mathbf{0}$ and $\mathrm{rank}\left(\mathbf{W}_{g}\right)=1$. In~\eqref{DelayAware35}, the rank one constraints are omitted. Similarly, problem~\eqref{DelayAware35} can be approximated as convex upper bound problem~\eqref{DelayAwarer45}, please see Appendix E for the details.
\begin{subequations}\label{DelayAwarer45}
\begin{align}
&\min~\eta+\theta+\frac{1}{2\rho}\sum\limits_{i\in\mathcal{K}_{\mathrm{R}}}
\mathds{1}\left(\sum\limits_{g\in\mathcal{G}}\overline{c}_{f_{g},i}\right)\left|\frac{S}{\theta}+
\phi\left(\mathbf{\Omega}_{i},\mathbf{\Omega}_{i}^{\left(t\right)}\right)-\ln\left(\left|\mathbf{A}_{i}
\right|\right)+\rho\lambda\right|^{2}\label{DelayAwarer45a}\\
\mathrm{s.t.}~&\eqref{DelayAwarer30b},\eqref{DelayAwarer30c},\eqref{DelayAwarer30d},\eqref{DelayAwarer30e},
\eqref{DelayAware35d},\label{DelayAwarer45b}\\
&\mathrm{r}_{g,1}-\ln\left(\mu_{k,1}\right)
+\phi\left(\chi_{k,1},\chi_{k,1}^{\left(t\right)}\right)\leq 0, \forall k\in\mathcal{K}_{\mathrm{U}},\label{DelayAwarer45c}\\
&S-\tau_{0}\mathrm{r}_{g,1}-\psi_{g}-\kappa_{g}\leq 0, \forall g\in\mathcal{G}\label{DelayAwarer45d}\\
&\phi\left(\tau_{0}+\theta,\tau_{0}+\theta^{\left(t\right)}\right)+\phi\left(\mathrm{r}_{g,1},
\mathrm{r}_{g,1}^{\left(t\right)}\right)-\ln\left(S\right)\leq 0\label{DelayAwarer45e}\\
&\phi\left(\psi_{g},\psi_{g}^{\left(t\right)}\right)-\ln\left(\theta\right)-\ln\left(\mathrm{r}_{g,1}\right)\leq 0, \forall g\in\mathcal{G}\label{DelayAwarer45f}\\
&\phi\left(\kappa_{g},\kappa_{g}^{\left(t\right)}\right)-\ln\left(\eta\right)-\ln\left(\mathrm{r}_{g,2}\right)\leq 0, \forall g\in\mathcal{G},\label{DelayAwarer45g}
\end{align}
\end{subequations}
where the optimization variables are $\eta$, $\theta$, $\kappa_{g}$, $\psi_{g}$, $\mathbf{W}_{g}$, $\overline{\mathbf{W}}_{g}$, $\mathbf{\Omega}$, $\mathrm{r}_{g,p}$, $\forall g\in\mathcal{G}$, $\forall p\in\mathcal{P}$. Note that in problem~\eqref{DelayAwarer45}, if the requested file $f_{g}$ is not stored at any eRRH, i.e., $\sum\limits_{i\in\mathcal{K}_{\mathrm{R}}}c_{f_{g},i}=0$, $r_{g,1}=0$ and constraints~\eqref{DelayAwarer45d},~\eqref{DelayAwarer45e},~\eqref{DelayAwarer45f}, and \eqref{DelayAwarer45g} are replaced with constraint~\eqref{DelayAwarer30f} for group $g$. If $S=\tau\mathrm{r}_{g,1}$ holds for group $g\in\mathcal{G}$, the constraint corresponding to group $g$ in~\eqref{DelayAwarer45d},~\eqref{DelayAwarer45f}, and~\eqref{DelayAwarer45g} are removed. It is not difficult to see that problem~\eqref{DelayAwarer45} is convex and can be solved with a classical convex optimization solver~\cite{MathPotra2000,AvialGrant2015}.

Our proposed algorithm for solving problem~\eqref{DelayAware35} consists of two loops. Specifically, we update the Lagrange multiplier $\lambda$ and the scalar penalty parameter $\rho$ according to certain criteria with fixed other optimization variables in the outer loop. While, in the inner loop, we use the classical optimization solver to solve problem~\eqref{DelayAwarer45}. The inner loop and the outer loop are alternative implemented until a certain stop criterion is satisfied. The detailed description for solving problem~\eqref{DelayAwarer45} is given in Algorithm~\ref{DelayAwareAlg03}, where $l$ and $t$ denotes the number of iterations, respectively, $\zeta^{\left(t\right)}$ represents the objective value of problem~\eqref{DelayAwarer45} at the $t$-th iteration and $0<\nu<1$. The analysis of the convergence and computational complexity is similar to that for Algorithm~\ref{DelayAwareAlg02} and is omitted here. In addition, the initialization of Algorithm~\ref{DelayAwareAlg03} can be realized using the method described in Appendix D.
\begin{algorithm}[htp]
\caption{Solution of problem~\eqref{DelayAware45}}\label{DelayAwareAlg03}
\begin{algorithmic}[1]
\STATE Set $\zeta^{\left(0\right)}$ to be a non-zero value and initialize non-zero beamforming matrix $\mathbf{W}_{g}^{\left(0\right)}$, $\overline{\mathbf{W}}_{g}^{\left(0\right)}$, $\forall g\in\mathcal{G}$, $\mathbf{\Omega}_{i}^{\left(0\right)}$, $\forall i\in\mathcal{K}_{\mathrm{R}}$, such that constraints~\eqref{DelayAware20c},~\eqref{DelayAware20d},~\eqref{DelayAware20e}, and~\eqref{DelayAware35d} are satisfied;\label{DelayAwareAlg0301}
\STATE Let $l=0$, initialize $\lambda^{\left(l\right)}$ and $\rho^{\left(l\right)}$ to be a non-zero value;\label{DelayAwareAlg0302}
\STATE Let $t=0$, and compute $\mu_{k,p}^{\left(t\right)}$ and $\chi_{k,p}^{\left(t\right)}$,  $\forall k\in\mathcal{K}_{\mathrm{U}}$, $\forall p\in\mathcal{P}$ with $\mathbf{W}_{g_{k}}^{\left(t\right)}$, $\overline{\mathbf{W}}_{g_{k}}^{\left(t\right)}$, and $\mathbf{\Omega}^{\left(t\right)}$. Let\label{DelayAwareAlg0303}
\begin{subequations}\label{DelayAware46}
\begin{align}
\theta^{\left(t\right)}&=\frac{S}{\min\limits_{i\in\mathcal{K}_{\mathrm{R}}}
\left(\ln\left(\left|\mathbf{A}_{i}^{\left(t\right)}
\right|\right)-\ln\left(\left|\mathbf{\Omega}_{i}^{\left(t\right)}\right|\right)\right)}
\label{DelayAware45a}\\
\mathrm{r}_{g,p}^{\left(t\right)}&=\nu\min\left(\min\limits_{k\in\mathcal{G}_{g}}
\frac{\ln\left(\mu_{k,p}^{\left(t\right)}\right)}{\ln\left(\chi_{k,p}^{\left(t\right)}\right)},\frac{S}{\tau_{0}
+\theta^{\left(t\right)}}\right)\label{DelayAware45b}\\
\psi_{g}^{\left(t\right)}&=\theta^{\left(t\right)}\mathrm{r}_{g,1}^{\left(t\right)}\label{DelayAware45c}\\
\kappa_{g}^{\left(t\right)}&=S-\left(\tau_{0}+\theta^{\left(t\right)}\right)\mathrm{r}_{g,1}^{\left(t\right)};\label{DelayAware45d}
\end{align}
\end{subequations}
\STATE Let $t\leftarrow t+1$. Solve problem~\eqref{DelayAware45} to obtain $\zeta^{\left(t\right)}$, $\eta^{\left(t\right)}$, $\theta^{\left(t\right)}$, $\kappa_{g}^{\left(t\right)}$, $\psi_{g}^{\left(t\right)}$, $\mathbf{W}_{g}^{\left(t\right)}$, $\overline{\mathbf{W}}_{g}^{\left(t\right)}$, $\forall g\in\mathcal{G}$, $\mathrm{r}_{g,p}^{\left(t\right)}$, $\forall p\in\mathcal{P}$, $\mathbf{\Omega}_{i}^{\left(t\right)}$, $\forall i\in\mathcal{K}_{\mathrm{R}}$;\label{DelayAwareAlg0304}
\STATE If $\left|\frac{\zeta^{\left(t\right)}-\zeta^{\left(t-1\right)}}{\zeta^{\left(t-1\right)}}\right|\leq\epsilon^{\left(l\right)}$, go to Step~\ref{DelayAwareAlg0306}. Otherwise, compute $\chi_{k,p}^{\left(t\right)}$, $\forall k\in\mathcal{K}_{\mathrm{U}}$, $\forall p\in\mathcal{P}$, and go to Step~\ref{DelayAwareAlg0304}; \label{DelayAwareAlg0305}
\STATE If $\left|\frac{S}{\theta^{\left(t\right)}}
-\min\limits_{i\in\mathcal{K}_{\mathrm{R}}}\left(\ln\left(\left|\mathbf{A}_{i}^{\left(t\right)}
\right|\right)-\ln\left(\left|\mathbf{\Omega}_{i}^{\left(t\right)}\right|\right)\right)\right|\leq\varepsilon$, stop iteration. Otherwise, go to Step~\ref{DelayAwareAlg0307};\label{DelayAwareAlg0306}
\STATE If $\left|\frac{S}{\theta^{\left(t\right)}}
-\min\limits_{i\in\mathcal{K}_{\mathrm{R}}}\left(\ln\left(\left|\mathbf{A}_{i}^{\left(t\right)}
\right|\right)-\ln\left(\left|\mathbf{\Omega}_{i}^{\left(t\right)}
\right|\right)\right)\right|\leq\varsigma^{\left(l\right)}$, update $\lambda$ and $\rho$ via~\eqref{DelayAware32}.\label{DelayAwareAlg0307}
Otherwise, update $\lambda$ and $\rho$ via~\eqref{DelayAware33};
\STATE Let $l\leftarrow l+1$, $\varsigma^{\left(l+1\right)}=\omega\left|\frac{S}{\theta^{\left(t\right)}}
-\min\limits_{i\in\mathcal{K}_{\mathrm{R}}}\left(\ln\left(\left|\mathbf{A}_{i}^{\left(t\right)}
\right|\right)-\ln\left(\left|\mathbf{\Omega}_{i}^{\left(t\right)}
\right|\right)\right)\right|$, and go to Step~\ref{DelayAwareAlg0303}.\label{DelayAwareAlg0308}
\end{algorithmic}
\end{algorithm}

\section*{\sc \uppercase\expandafter{\romannumeral5}. Numerical Results}
\begin{figure}[t]
\renewcommand{\captionfont}{\footnotesize}
\renewcommand*\captionlabeldelim{.}
\centering
\captionstyle{flushleft}
\onelinecaptionstrue
\includegraphics[width=0.6\columnwidth,keepaspectratio]{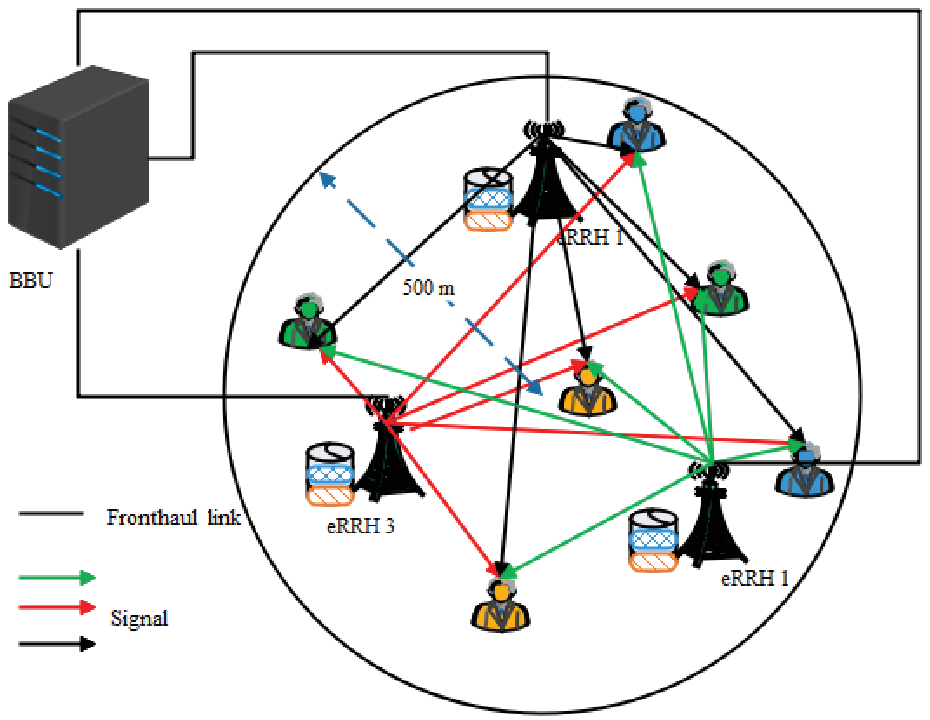}\\
\caption{Simulation Model, $K_{\mathrm{R}}=3$ and $K_{\mathrm{U}}=6$.}
\label{SimulationModel}
\end{figure}

In this section, we present numerical results to evaluate the performance of the proposed transmission schemes for cache-enabled multigroup multicasting RANs. For simplicity, we consider that all eRRHs have the same maximum transmit power and fronthaul capacity, i.e., $P_{i}=P$ and $C_{i}=C$, $\forall i\in\mathcal{K}_{\mathrm{R}}$. In the cache-enabled multigroup multicasting RAN system, the positions of eRRHs and users are uniformly distributed within a circular cell of radius $500$ m, as illustrated in Fig.~\ref{SimulationModel}. The channel vector $\mathbf{h}_{k,i}$ from eRRH $i$ to user $k$ is modeled as $\mathbf{h}_{k,i}=\sqrt{\varrho_{k,i}}\widetilde{\mathbf{h}}_{k,i}$ , where the channel power $\varrho_{k,i}$ is given as $\varrho_{k,i} = 1/\left(1 + (d_{k,i}/d_{0}\right)^{\alpha})$ and the elements of $\widetilde{\mathbf{h}}_{k,i}$ are independent and identically distributed (i.i.d.) with $\mathcal{CN}\left(0,1\right)$. All users have the same noise variance, i.e., $\sigma_{k}^{2}=\sigma^{2}$, $k\in\mathcal{K}_{\mathrm{U}}$. The eRRHs are equipped with caches of equal size, i.e., $B_{i}=B=\lfloor\xi SF\rfloor$, $i\in\mathcal{K}_{\mathrm{R}}$, where $\xi$ denotes the fractional caching proportion. The cache states $c_{f,i}$, $f\in\mathcal{F}$, $i\in\mathcal{K}_{\mathrm{R}}$, are randomly generated. To illustrate the effectiveness of the proposed schemes, we include the numerical performance of the transmission scheme which aims to maximize the minimum delivery rate, labeled as ``JCEO Scheme"~\cite{TWCPark2016}. If not stated otherwise, the simulation is performed with the parameters given in Table~\ref{SimulatedParametersValues}. 

\begin{table}[t]
	\setlength{\abovecaptionskip}{0pt}
	\setlength{\belowcaptionskip}{5pt}
	\captionstyle{flushleft}
	\onelinecaptionstrue
	\centering
	\caption{Simulation parameters}
	\begin{tabular}{|c|c||c|c||c|c|}
		\hline
		Symbol&\makecell[c]{Value}&Symbol&\makecell[c]{Value}&Symbol&\makecell[c]{Value}\\
		\hline
		$K_{\mathrm{R}}$& $3$&$K_{\mathrm{U}}$ &$3/6$&$N_{\mathrm{t}}$& $1/4$\\
        \hline
		$\xi$ & $0.5$&$\sigma^{2}$ &$1$&$d_{0}$ &$50$ m\\
		\hline
         $F$ & $10$	&$\alpha$ &$3$& $\tau_{0}$ &$10$ ms\\
		\hline
         $\varepsilon$& $10^{-5}$&$\varsigma^{\left(0\right)}$&$10^{-3}$& $\epsilon^{\left(0\right)}$ &$10^{-3}$\\
		\hline
         $\lambda^{\left(0\right)}$& $0.5$&$\rho^{\left(0\right)}$&$0.5$& $\omega$ &$0.6$\\
		\hline
        $\nu$& $0.1$&$\delta$&$0.5$& $-$ &$-$\\
		\hline
	\end{tabular}
	\label{SimulatedParametersValues}
\end{table}

\begin{figure}[t]
\renewcommand{\captionfont}{\footnotesize}
\renewcommand*\captionlabeldelim{.}
\centering
\captionstyle{flushleft}
\onelinecaptionstrue
\includegraphics[width=0.8\columnwidth,keepaspectratio]{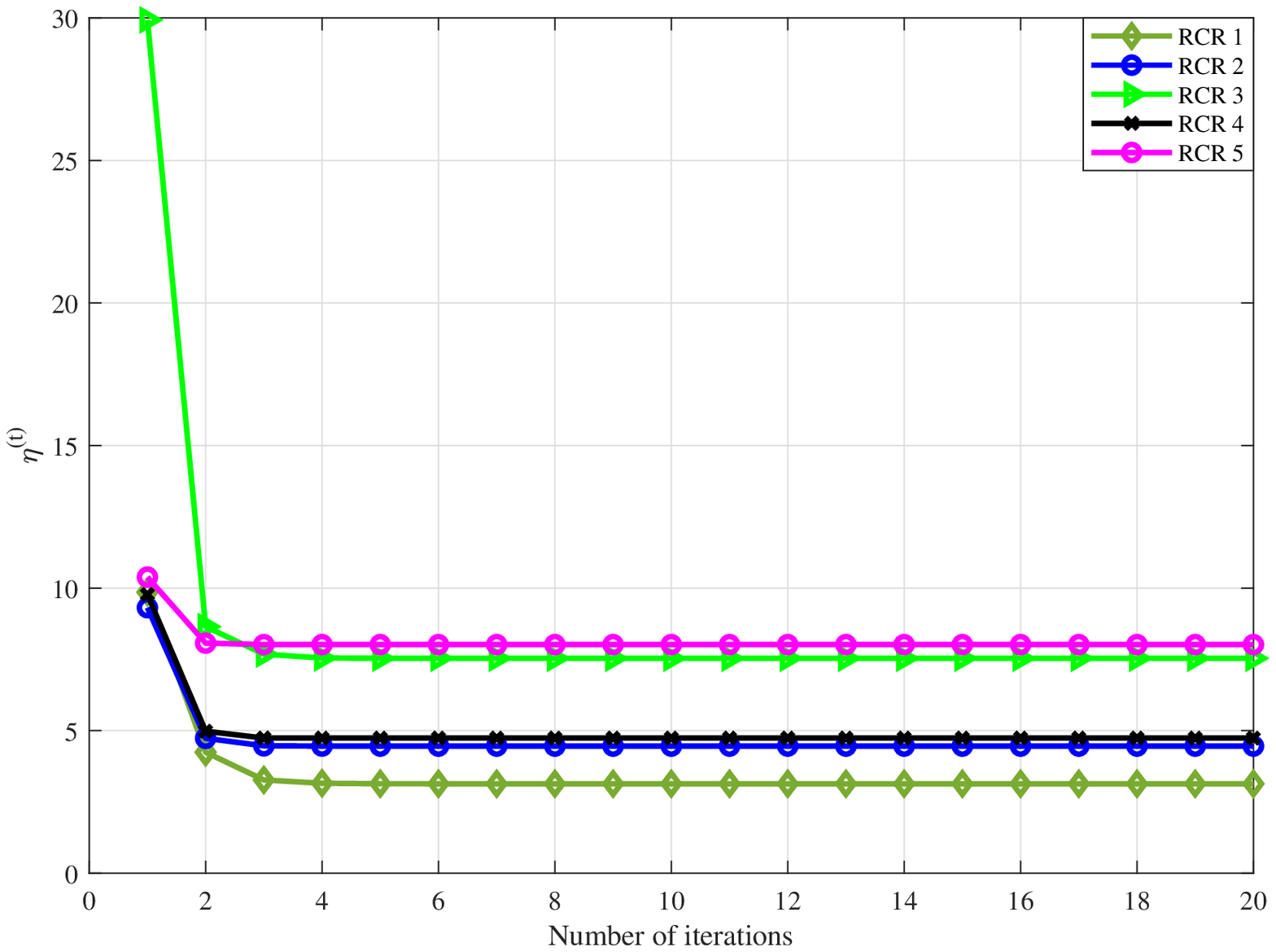}\\
\caption{Convergence trajectories of Algorithm~\ref{DelayAwareAlg01} for different RCRs.}
\label{FullCachingConvergenceTrajectory}
\end{figure}

\begin{figure}[t]
\renewcommand{\captionfont}{\footnotesize}
\renewcommand*\captionlabeldelim{.}
\centering
\captionstyle{flushleft}
\onelinecaptionstrue
\includegraphics[width=0.8\columnwidth,keepaspectratio]{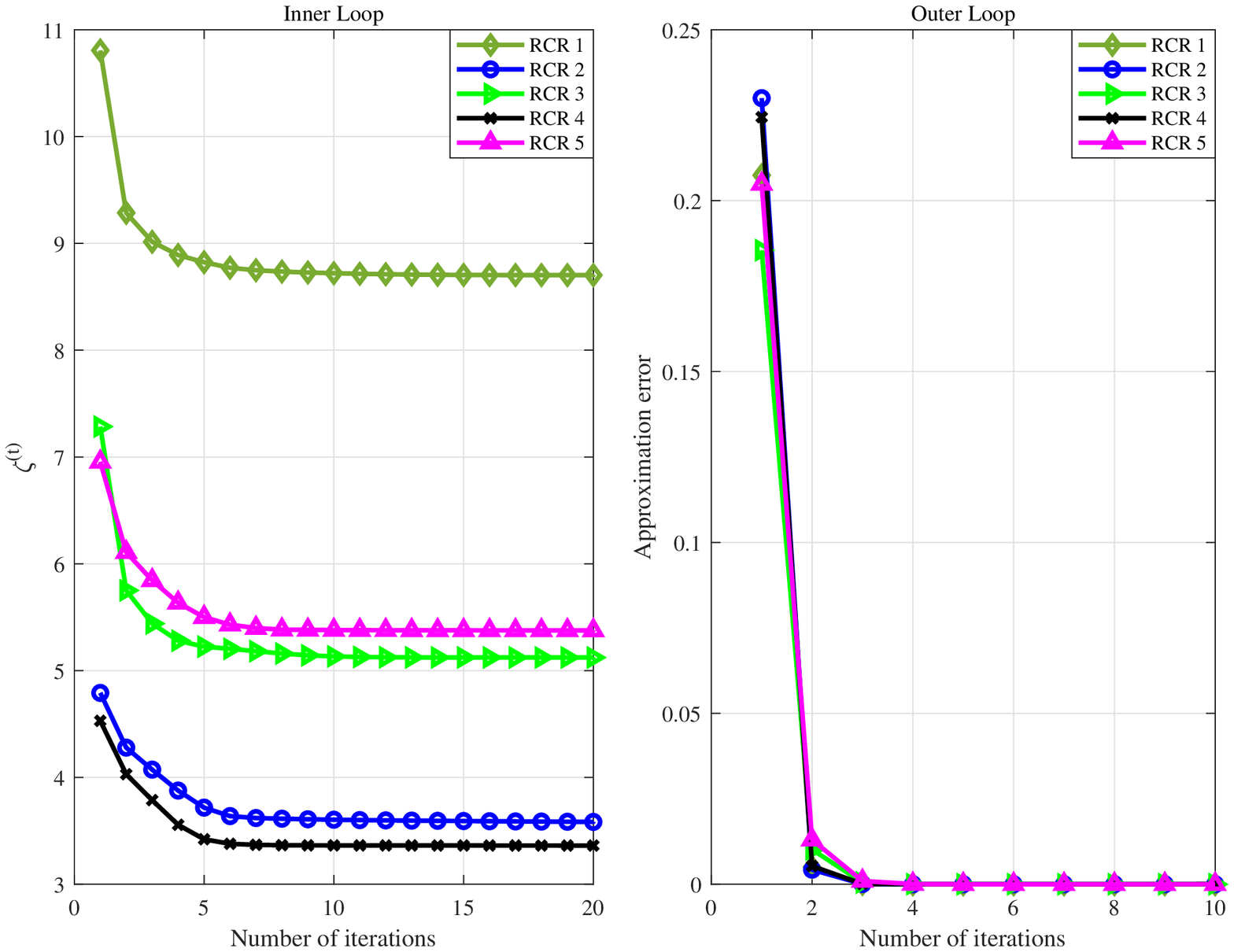}\\
\caption{Convergence trajectories of Algorithm~\ref{DelayAwareAlg03} for different RCRs.}
\label{TwophaseConvergenceTrajectory}
\end{figure}

Fig.~\ref{FullCachingConvergenceTrajectory} and Fig.~\ref{TwophaseConvergenceTrajectory} illustrate the convergence trajectory of Algorithm~\ref{DelayAwareAlg01} and Algorithm~\ref{DelayAwareAlg03} for different random channel realization (RCR) with $S=1.5$ nats/Hz, $P=20$ dB, $C=2$ nats/Hz/s, $K_{\mathrm{U}}=3$, and $N_{\mathrm{t}}=1$. In the right subfigure of Fig.~\ref{TwophaseConvergenceTrajectory}, the approximation error is defined as
$\left|\theta-\frac{S}{\min\limits_{i\in\mathcal{K}_{\mathrm{R}}}\left(\ln\left(\left|\mathbf{A}_{i}
\right|\right)-\ln\left(\left|\mathbf{\Omega}_{i}\right|\right)\right)}\right|$. Fig.~\ref{FullCachingConvergenceTrajectory} demonstrates that a non-increasing sequence is generated with the running of Algorithm~\ref{DelayAwareAlg01}. The inner loop and the outer loop of Algorithm~\ref{DelayAwareAlg03} also generate a non-increasing sequence, respectively, as illustrated in Fig.~\ref{TwophaseConvergenceTrajectory}. Recalling the bounded properties of the objective of problems~\eqref{DelayAware18} and~\eqref{DelayAware44}, the convergence of Algorithm~\ref{DelayAwareAlg01} and Algorithm~\ref{DelayAwareAlg03} can be guaranteed~\cite{OperalMarks1977}\footnote{As Algorithm~\ref{DelayAwareAlg02} is similar to Algorithm~\ref{DelayAwareAlg03}, we only present the convergence trajectory of Algorithm~\ref{DelayAwareAlg03}. The convergence of Algorithm~\ref{DelayAwareAlg02} can be guaranteed as well.}.

\begin{figure}[ht]
\renewcommand{\captionfont}{\footnotesize}
\renewcommand*\captionlabeldelim{.}
\centering
\captionstyle{flushleft}
\onelinecaptionstrue
\includegraphics[width=0.8\columnwidth,keepaspectratio]{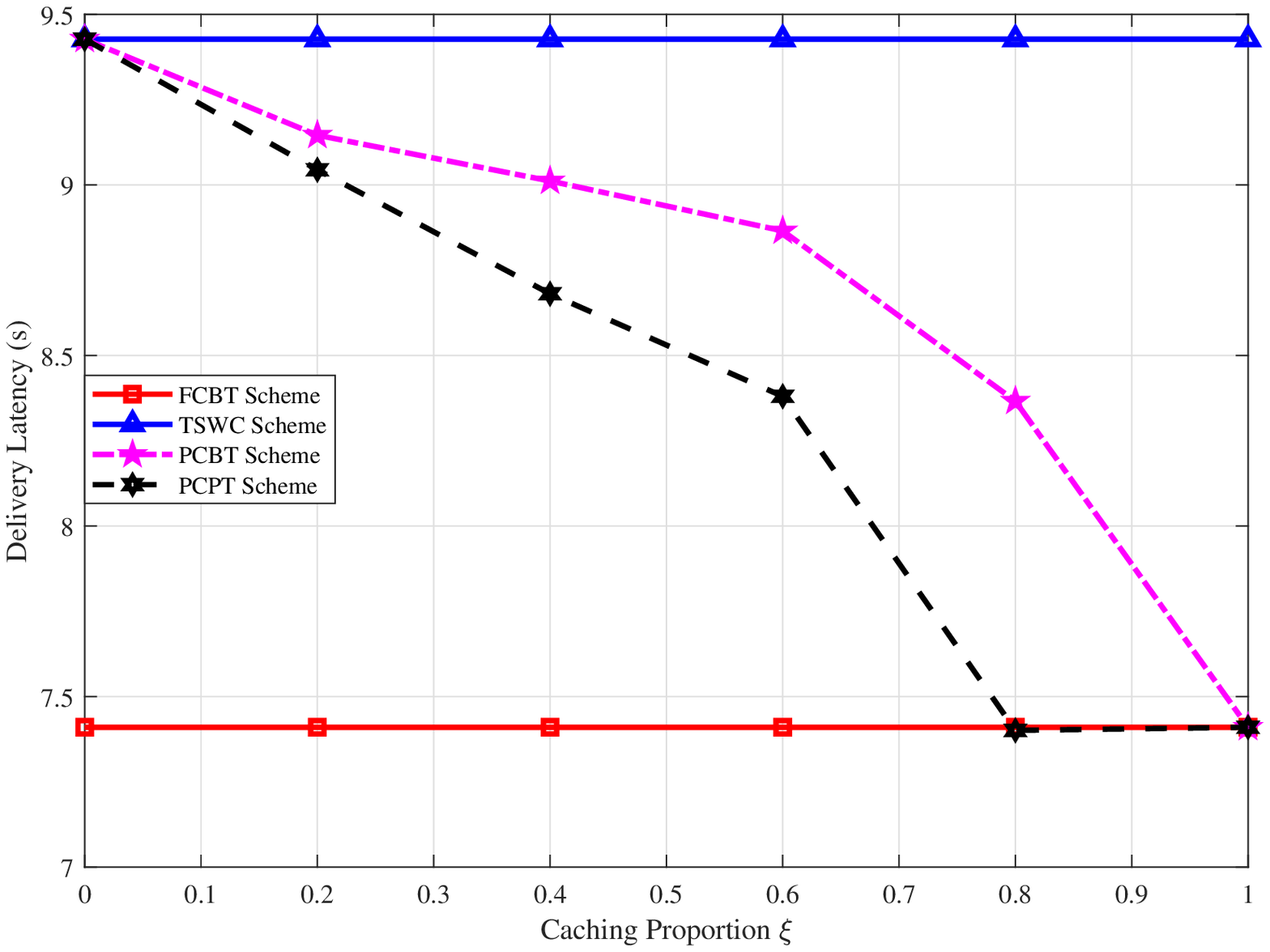}\\
\caption{Delivery latency versus the caching proportion $\xi$.}
\label{LatencyVsCachingProportion}
\end{figure}

Fig.~\ref{LatencyVsCachingProportion} shows the delivery latency versus the fractional caching proportion $\xi$ with $S=1.5$ nats/Hz, $P=20$ dB, $C=2$ nats/Hz/s, $K_{\mathrm{U}}=6$, $G=3$, and $N_{\mathrm{t}}=1$. It can be observed that the FCBT scheme achieves the best performance in terms of delivery latency. The transmission scheme without caching (TSWC) scheme achieves the largest delivery latency. This is because all requested files need to be fetched from the BBU such that the limited capacity of fronthaul links has significant negative impact on the system performance. The delivery latency of the other two transmission schemes decreases as the fractional caching proportion $\xi$ increases. The larger the fractional caching proportion $\xi$, the greater the probability of the requested files that are cached at the local cache. Thus, the impact of the capacity of fronthaul links on the system performance is reduced. In addition, the PCPT scheme outperforms the PCBT scheme, because the PCPT scheme takes advantage of the delay $\tau$ interval to transmit cached requested files before the uncached requested files arrive at the eRRHs.

\begin{figure}[t]
\renewcommand{\captionfont}{\footnotesize}
\renewcommand*\captionlabeldelim{.}
\centering
\captionstyle{flushleft}
\onelinecaptionstrue
\includegraphics[width=0.8\columnwidth,keepaspectratio]{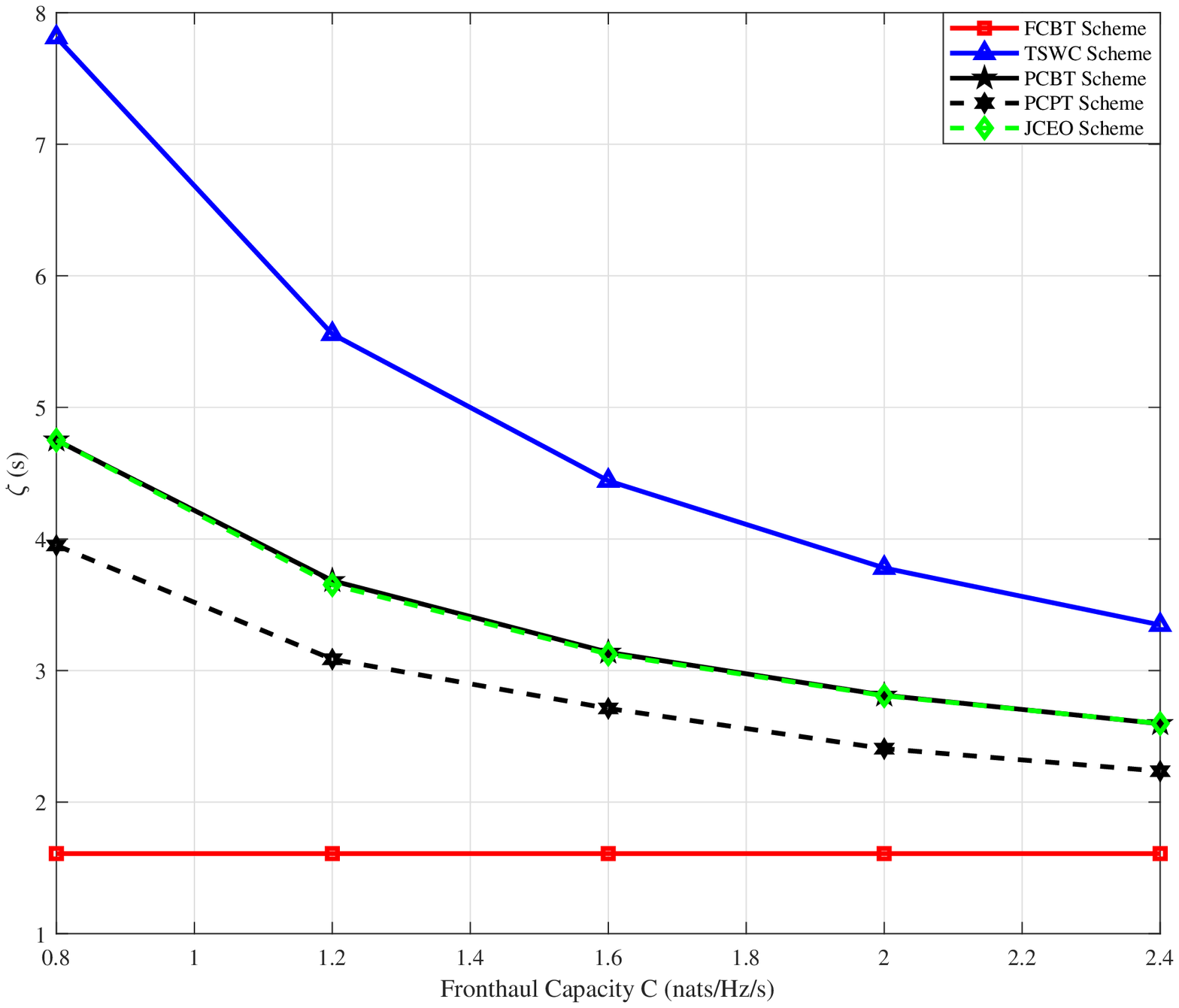}\\
\caption{Delivery latency versus the fronthaul capacity $C$.}
\label{LatencyVsCapacity}
\end{figure}

Fig.~\ref{LatencyVsCapacity} illustrates the delivery latency versus the fronthaul capacity $C$ with $S=1.2$ nats/Hz, $P=20$ dB, $K_{\mathrm{U}}=6$, and $N_{\mathrm{t}}=4$. Except for the FCBT scheme that is not limited by the capacity $C$ of the fronthaul links, the delivery latency achieved by the other three transmission schemes decreases with an increasing capacity $C$ of fronthaul links. This implies that caching at the network edge helps to reduce the burden on the fronthaul links, i.e., the amount of requested files being fetched from the BBS decreases. Therefore, the impact of constraints~\eqref{DelayAware10d} on the performance of the PCBT and PCPT schemes decreases. In addition, the second item of~\eqref{DelayAware11} may reduce with fronthaul capacity $C$ increases. As a consequence, the delivery latency reduces. When the system performance is limited to fronthaul capacity $C$ or the achievable rate $R_{k}$, $k\in\mathcal{K}_{\mathrm{U}}$, i.e., is not limited to file size $S$, the PCBT and JCEO schemes achieve the same delivery latency. This is because they make full use of all resources to maximize the minimum delivery rate and the rate on the fronthaul links, i.e., minimize the delivery latency defined in~\eqref{DelayAware10a} which takes into account the fairness for all multicast groups. Compared to the TSWC, PCBT, and JCEO schemes, the performance achieved by the PCPT scheme is closer to that of the FCBT scheme. Except for exploiting delay $\tau$ to transmit requested files, an advantage of the PCPT scheme is to increase the degree of freedom for power allocation in each transmission phase and to reduce the inter-group interference for cache-enabled multigroup multicasting RANs. As a consequence, the system throughput can be increased and the delivery latency is reduced.

\begin{figure}[t]
\renewcommand{\captionfont}{\footnotesize}
\renewcommand*\captionlabeldelim{.}
\centering
\captionstyle{flushleft}
\onelinecaptionstrue
\includegraphics[width=0.8\columnwidth,keepaspectratio]{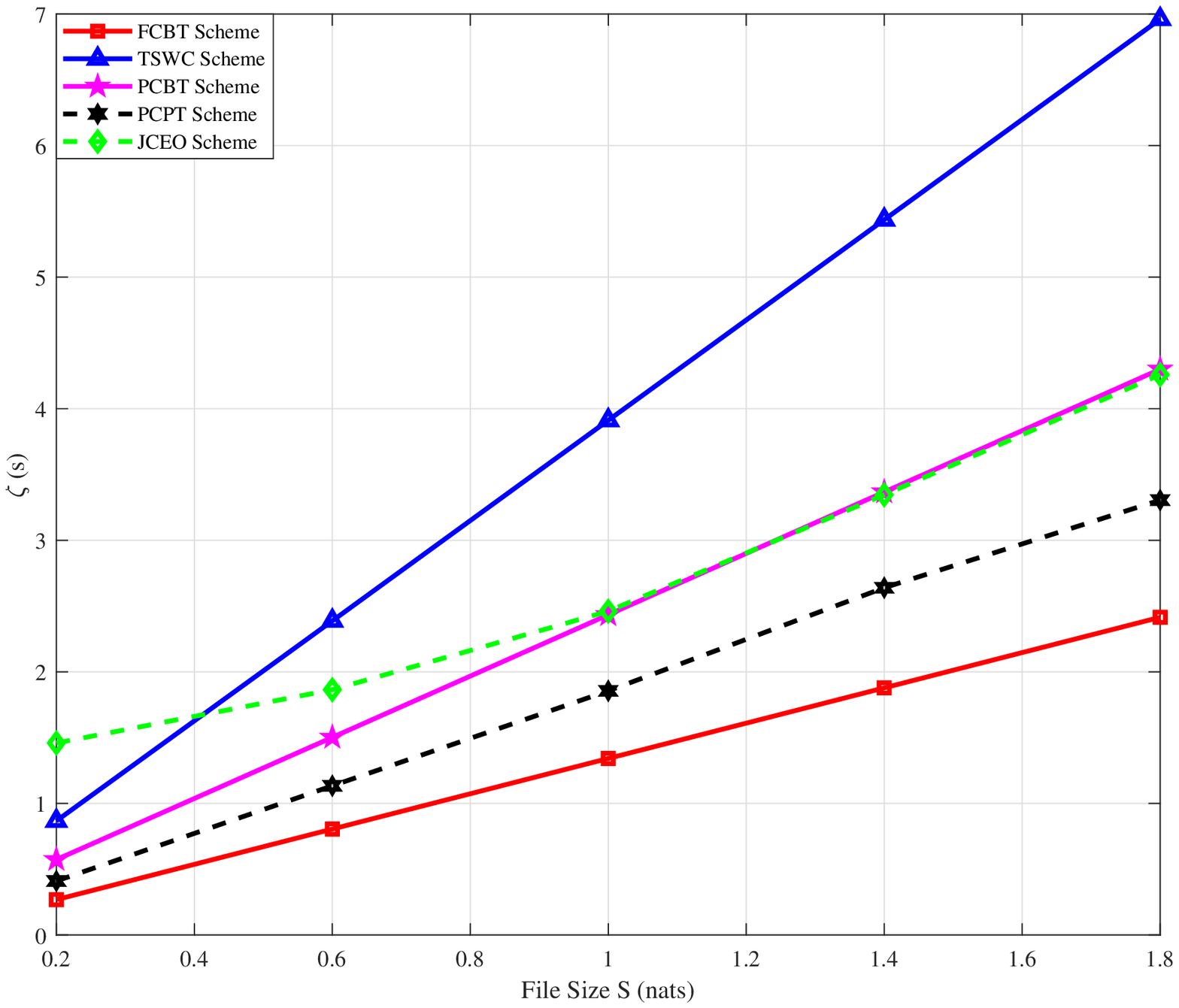}\\
\caption{Delivery latency versus the file size $S$.}
\label{LatencyVsFileSize}
\end{figure}

Fig.~\ref{LatencyVsFileSize} shows the delivery latency versus file size $S$ with $C=1.5$ nats/Hz/s, $P=20$ dB, $K_{\mathrm{U}}=6$, and $N_{\mathrm{t}}=4$. Given the channel statistics and the capacity $C$ of the fronthaul links, the time to transmit all requested files increases as file size $S$ increases from the objective function of problems~\eqref{DelayAware09},~\eqref{DelayAware10}, and~\eqref{DelayAware11}, respectively. At the same time, the burden on the fronthaul links also increases with increasing file size $S$. Results illustrated in Fig.~\ref{LatencyVsFileSize} demonstrate that the delivery latency of all transmission schemes increases as file size $S$ increases. Compared to the other non-FCBT transmission schemes, the PCPT scheme obtains the minimum delivery latency, since it effectively exploits the delay interval incurred by fetching the uncached requested files from the BBU and by the baseband signal processing at the BBU to transmit requested files. When file size $S$ is smaller, the proposed PCBT and PCPT schemes outperform the JCEO scheme in terms of delivery latency. This is because the delivery rate of the JCEO scheme is limited by the file size. However, when the system performance is not limited to file size $S$, i.e., file size $S$ is sufficiently large, the PCBT and JCEO schemes achieve the same delivery latency since they make full use of all resources to maximize the minimum delivery rate and the rate on the fronthaul links. It also means that the PCBT and JCEO schemes guarantee the fairness among all multicast groups. The delivery latency of the TSWC scheme increases rapidly as file size $S$ increases as the fronthaul links become more congested. The results indicate that the network edge caching becomes more and more important as file size $S$ increases, especially for the delay-sensitive data traffic.

\section*{\sc \uppercase\expandafter{\romannumeral6}. Conclusions}

In this paper, according to the caching capability of each eRRH, we investigated three transmission schemes to minimize the delivery latency for cache-enabled multigroup multicasting RANs. Correspondingly, three delivery latency minimization problems were formulated. The formulated optimization problems are non-convex and difficult to obtain directly the global optimum solutions. We further developed an efficient algorithm to address each delivery latency minimization problem. Finally, numerical results were provided to valuate the effectiveness of the proposed algorithms and shown that the PCPT scheme outperforms the PCBT scheme in terms of delivery latency. It can be observed that if the transmission scheme is designed carefully, caching frequently requested content at the network edge can effectively reduce the delivery latency while reducing the burden on the fronthaul links.

\section*{Appendix}

\subsection*{A. Convergence and Computational Complexity Analysis}

\emph{1) Convergence analysis of Algorithm~\ref{DelayAwareAlg01}:} Consider the $\left(t+1\right)$-th iteration of Algorithm~\ref{DelayAwareAlg01} that solves the optimization problem~\eqref{DelayAware18}. If we replace $\eta$, $\mathbf{w}_{g}$, $\mathrm{r}_{g,1}$, $\forall g\in\mathcal{G}$, $\overline{\gamma}_{k,1}$ and $\chi_{k,1}$, $\forall k\in\mathcal{K}_{\mathrm{U}}$ with $\eta^{\left(t\right)}$, $\mathbf{w}_{g}^{\left(t\right)}$, $\mathrm{r}_{g,1}^{\left(t\right)}$, $\forall g\in\mathcal{G}$, $\overline{\gamma}_{k,1}^{\left(t\right)}$ and $\chi_{k,1}^{\left(t\right)}$, $\forall k\in\mathcal{K}_{\mathrm{U}}$, respectively, all constraints are still satisfied, which means that the solution of the $t$-th iteration is feasible point of problem~\eqref{DelayAware18} in the $\left(t+1\right)$-th iteration. Thus, the objective function obtained in the $\left(t+1\right)$-th iteration is not larger than that in the $t$-th iteration, i.e., $\eta^{\left(t+1\right)}\leq \eta^{\left(t\right)}$. That is to say, Algorithm~\ref{DelayAwareAlg01} generates a non-increasing sequence of objective value $\eta^{\left(t\right)}$. Moreover, the problem is bounded due to the power constraints. Therefore, the convergence of Algorithm~\ref{DelayAwareAlg01} can be guaranteed by the monotonic boundary theorem~\cite{Bibby1974}. Furthermore, based on the arguments presented in~\cite[Theorem 1]{OperalMarks1977}, we can prove that Algorithm~\ref{DelayAwareAlg01} converges to a KKT solution of problem~\eqref{DelayAware10}.

\emph{2) Computational complexity analysis of Algorithm~\ref{DelayAwareAlg01}:}In Algorithm~\ref{DelayAwareAlg01}, Step~\ref{DelayAwareAlg0103} solves a convex problem, which can be efficiently implemented by the primal-dual interior point method with approximate complexity of $\mathit{O}\left(\left(G\left(N_{\mathrm{t}}K_{\mathrm{R}}+4\right)\right)^{3.5}\right)$, where  $\mathit{O}\left(\cdot\right)$ stands for the big-O notation~\cite{MathPotra2000}. Letting $\upsilon_{1}$ be the number of iterations in Algorithm~\ref{DelayAwareAlg01}, the overall computational complexity is $\mathit{O}\left(\upsilon_{1}\left(G\left(N_{\mathrm{t}}K_{\mathrm{R}}+4\right)\right)^{3.5}\right)$.

\subsection*{B. Convexity of Problem~\eqref{DelayAwarer30}}

In this subsection, we address the difficulties of solving problem~\eqref{DelayAware20} step-by-step. First, we convexify non-convex constraints~\eqref{DelayAware20b} and~\eqref{DelayAware20e}. Then, we convexify objective function~\eqref{DelayAware20a}. According to the concave property of function $\ln\left(\cdot\right)$, constraint~\eqref{DelayAware20b} can be convexified as
\begin{equation}\label{DelayAware23}
\mathrm{r}_{g,2}-\ln\left(\mu_{k,2}\right)+\phi\left(\chi_{k,2},\chi_{k,2}^{\left(t\right)}\right)
\leq 0, \forall k\in\mathcal{K}_{\mathrm{U}}.
\end{equation}
Similarly, constraint~\eqref{DelayAware20e} can be approximated with the following convex form
\begin{equation}\label{DelayAware24}
\phi\left(\mathbf{A}_{i},\mathbf{A}_{i}^{\left(t\right)}\right)
-\ln\left(\left|\mathbf{\Omega}_{i}\right|\right)\leqslant C_{i}, \forall i\in\mathcal{K}_{\mathrm{R}}.
\end{equation}
In~\eqref{DelayAware24}, $\mathbf{A}_{i}$ is redefined as $\mathbf{A}_{i}=\sum\limits_{g\in\mathcal{G}}\mathbf{P}_{g,i}^{T}\left(\overline{c}_{f_{g},i}\right)\overline{\mathbf{W}}_{g}
\mathbf{P}_{g,i}\left(\overline{c}_{f_{g},i}\right)+\mathbf{\Omega}_{i}$,
In~\eqref{DelayAware23} and~\eqref{DelayAware24}, $\phi\left(\mathbf{A}, \mathbf{B}\right)$ is defined as
\begin{equation}\label{DelayAware25}
\phi\left(\mathbf{A}, \mathbf{B}\right)=\ln\left(\left|\mathbf{B}\right|\right)+
\mathrm{tr}\left(\mathbf{B}^{-1}\left(\mathbf{A}-\mathbf{B}\right)\right).
\end{equation}
In problem~\eqref{DelayAware20}, if we replace constraints~\eqref{DelayAware20b} and~\eqref{DelayAware20e} with~\eqref{DelayAware23} and~\eqref{DelayAware24}, respectively, all constraints in~\eqref{DelayAware20} are transformed into convex forms. Thus, problem~\eqref{DelayAwarer30} can be reformulated as
\begin{subequations}\label{DelayAware26}
\begin{align}
&\min~\eta+\theta\label{DelayAware26a}\\
\mathrm{s.t.}~&\eqref{DelayAware20c},\eqref{DelayAware20d}, \eqref{DelayAware23}, \eqref{DelayAware24}\label{DelayAware26b}\\
&\ln\left(S\right)-\ln\left(\eta\right)-\ln\left(\mathrm{r}_{g,2}\right)\leq 0, \forall g\in\mathcal{G} \label{DelayAware26c}\\
&\theta=\frac{S}{\min\limits_{i\in\mathcal{K}_{\mathrm{R}}}\left(\ln\left(\left|\mathbf{A}_{i}
\right|\right)-\ln\left(\left|\mathbf{\Omega}_{i}\right|\right)\right)} \label{DelayAware26d}
\end{align}
\end{subequations}
where the optimization variables are $\eta$, $\theta$, $\overline{\mathbf{W}}_{g}$, $\mathbf{\Omega}$, $\mathrm{r}_{g,2}$, $\forall g\in\mathcal{G}$. Note that constant $\tau_{0}$ in objective function~\eqref{DelayAware26a} is omitted. In constraint~\eqref{DelayAware26c}, we exploit the positive nature of $\eta$ and $r_{g,2}$, i.e., $\eta>0$ and $r_{g,2}>0$. This is because if $r_{g,2}=0$, the delivery latency is infinite, i.e., problem~\eqref{DelayAware20} becomes meaningless. The new challenge of solving problem~\eqref{DelayAware26} is the introduction of equality constraint~\eqref{DelayAware26d} which is non-convex. To remove the equality constraint, we use the Lagrangian duality method to address problem~\eqref{DelayAware26}. The partial augmented Lagrangian function of problem~\eqref{DelayAware26} is given by
\begin{subequations}\label{DelayAware27}
\begin{align}
&\min~\eta+\theta+\frac{1}{2\rho}\left|\frac{S}{\theta}-\min\limits_{i\in\mathcal{K}_{\mathrm{R}}}
\left(\ln\left(\left|\mathbf{A}_{i}
\right|\right)-\ln\left(\left|\mathbf{\Omega}_{i}\right|\right)\right)
+\rho\lambda\right|^{2}\label{DelayAware27a}\\
\mathrm{s.t.}~&\eqref{DelayAware20c}, \eqref{DelayAware20d}, \eqref{DelayAware23}, \eqref{DelayAware24},\eqref{DelayAware26c} \label{DelayAware27b}
\end{align}
\end{subequations}
where the optimization variables are $\eta$, $\theta$, $\overline{\mathbf{W}}_{g}$, $\mathbf{\Omega}$, $\mathrm{r}_{g,2}$, $\forall g\in\mathcal{G}$. In problem~\eqref{DelayAware27}, $\lambda$ is the Lagrange multiplier and $\rho$ is a scalar penalty parameter. This penalty parameter improves the robustness compared to other optimization methods for constrained problems (e.g. dual ascent method) and in particular achieves convergence without the need of specific assumptions for the objective function, i.e. strict convexity and finiteness~\cite{FoundBoyd2011,JSACTsinos2017,JSTSPShi2018,TSPSun2018}. The smaller the value of $\rho$, the greater the probability of equality~\eqref{DelayAware26d} holds.

It is still challenging to address problem~\eqref{DelayAware27} directly, since the third term in objective function~\eqref{DelayAware27a} is difficult to tackle. To further obtain a tractable form of problem~\eqref{DelayAware27}, we relax problem~\eqref{DelayAware27} to the following form
\begin{subequations}\label{DelayAware28}
\begin{align}
&\min~\eta+\theta+\frac{1}{2\rho}\sum\limits_{i\in\mathcal{K}_{\mathrm{R}}}
\mathds{1}\left(\sum\limits_{g\in\mathcal{G}}\overline{c}_{f_{g},i}\right)\left|\frac{S}{\theta}+
\ln\left(\left|\mathbf{\Omega}_{i}\right|\right)-\ln\left(\left|\mathbf{A}_{i}
\right|\right)+\rho\lambda\right|^{2}\label{DelayAware28a}\\
\mathrm{s.t.}~&\eqref{DelayAware20c}, \eqref{DelayAware20d}, \eqref{DelayAware23}, \eqref{DelayAware24},\eqref{DelayAware26c}, \label{DelayAware28b}
\end{align}
\end{subequations}
where the optimization variables are $\eta$, $\theta$, $\overline{\mathbf{W}}_{g}$, $\mathbf{\Omega}$, $\mathrm{r}_{g,2}$, $\forall g\in\mathcal{G}$. In~\eqref{DelayAware28}, indicator function $\mathds{1}\left(x\right)$ is defined as follows:
\begin{equation}\label{DelayAware29}
\mathds{1}\left(x\right)=
\begin{cases}
0, & \mathrm{if}~x= 0, \\
1, & \mathrm{if}~x\neq 0.
\end{cases}
\end{equation}
When all requested files are stored at eRRH $i$, the value of $\ln\left(\left|\mathbf{A}_{i}\right|\right)-\ln\left(\left|\mathbf{\Omega}_{i}\right|\right)$ should be a very large constant value. Therefore, without affecting the solution of~\eqref{DelayAware28}, in~\eqref{DelayAware28a}, the indication function $\mathds{1}\left(x\right)$ is introduced to move away from the item related to eRRH $i$ which stores all requested files at its local cache. At the optimal point of problem~\eqref{DelayAware28}, there is at least an eRRH $i$ such that equality constraint~\eqref{DelayAware26d} holds. Using again the SCA method, problem~\eqref{DelayAware28} can be further convexified as the following convex upper bound problem
\begin{subequations}\label{DelayAware30}
\begin{align}
&\min~\eta+\theta+\frac{1}{2\rho}\sum\limits_{i\in\mathcal{K}_{\mathrm{R}}}
\mathds{1}\left(\sum\limits_{g\in\mathcal{G}}\overline{c}_{f_{g},i}\right)\left|\frac{S}{\theta}+
\phi\left(\mathbf{\Omega}_{i},\mathbf{\Omega}_{i}^{\left(t\right)}\right)-
\ln\left(\left|\mathbf{A}_{i}\right|\right)+\rho\lambda\right|^{2}\label{DelayAware30a}\\
\mathrm{s.t.}~&\eqref{DelayAware20c}, \eqref{DelayAware20d}, \eqref{DelayAware23}, \eqref{DelayAware24},\eqref{DelayAware26c}, \label{DelayAware30b}
\end{align}
\end{subequations}
where the optimization variables are $\eta$, $\theta$, $\overline{\mathbf{W}}_{g}$, $\mathbf{\Omega}$, $\mathrm{r}_{g,2}$, $\forall g\in\mathcal{G}$.

\subsection*{C. Gaussian Randomization}

Due to the rank relaxation, in general, the solution to problem~\eqref{DelayAwarer30}, denoted as $\eta^{\left(o\right)}$, $\theta^{\left(o\right)}$, $\overline{\mathbf{W}}_{g}^{\left(o\right)}$, $\mathbf{\Omega}^{\left(o\right)}$, $\mathrm{r}_{g,2}^{\left(o\right)}$, $\forall g\in\mathcal{G}$, may not comprise only rank-one matrices $\overline{\mathbf{W}}_{g}^{\left(o\right)}$, $\forall g\in\mathcal{G}$. Hence, the optimum beamforming vectors cannot be directly extracted from the obtained $\overline{\mathbf{W}}_{g}^{\left(o\right)}$. If $\overline{\mathbf{W}}_{g}^{\left(o\right)}$, $\forall g\in\mathcal{G}$, is of rank one, we can write $\overline{\mathbf{W}}_{g}^{\left(o\right)}=\overline{\mathbf{w}}_{g}^{\left(o\right)}
\overline{\mathbf{w}}_{g}^{\left(o\right)H}$ and $\overline{\mathbf{w}}_{g}^{\left(o\right)}$ will be a feasible solution to problem~\eqref{DelayAware20}. If the rank of $\overline{\mathbf{W}}_{g}^{\left(o\right)}$ is larger than $1$, we can use the Gaussian randomization technique to generate candidate beamformer vector from~$\overline{\mathbf{W}}_{g}^{\left(o\right)}$. In the randomization technique, we eigendecompose $\overline{\mathbf{W}}_{g}^{\left(o\right)}=\overline{\mathbf{U}}_{g}\mathbf{\Lambda}_{g}\overline{\mathbf{U}}_{g}^{H}$ and choose $\overline{\mathbf{w}}_{g}^{\left(o\right)}=\sqrt{p_{g}}
\overline{\mathbf{U}}_{g}\mathbf{\Lambda}_{g}^{1/2}\mathbf{e}_{g}$, where $\mathbf{e}_{g}\sim\mathcal{CN}\left(\mathbf{0},\mathbf{I}\right)$ and $p_{g}$ denotes the sought power boost (or reduction) factor for multicast group $g$~\cite{SIAMVand1996,TSPSidi2006,TSPKari2008,TSPTervo2018}. The specific value of $p_{g}$, $\forall g\in\mathcal{G}$, can be obtained by solving the following problem
\begin{subequations}\label{DelayAware34}
\begin{align}
&\min\sum\limits_{g\in\mathcal{G}}p_{g}\label{DelayAware34a}\\
\mathrm{s.t.}~&\mathrm{r}_{{g},2}^{\left(o\right)}\leq\mathrm{R}_{k,2}^{\left(o\right)}, \forall k\in\mathcal{G}_{g}, \forall g\in\mathcal{G} \label{DelayAware34b}\\
&\sum\limits_{g\in\mathcal{G}}\left\|\mathbf{P}_{g,i}^{T}\left(1\right)\overline{\mathbf{w}}_{g}^{\left(o\right)}\right\|^{2}+
\mathrm{Tr}\left(\mathbf{\Omega}_{i}^{\left(o\right)}\right)\leq P_{i}, \forall i\in\mathcal{K}_{\mathrm{R}}\label{DelayAware34c}\\
&g_{i}\left(\mathcal{V}^{\left(o\right)},\mathcal{O}^{\left(o\right)}\right)\leqslant C_{i}, \forall i\in\mathcal{K}_{\mathrm{R}}\label{DelayAware34d}
\end{align}
\end{subequations}
where the optimization variables are $p_{g}$, $\forall g\in\mathcal{G}$. In~\eqref{DelayAware34b}, $\mathrm{R}_{k,2}^{\left(o\right)}=\ln\left(1+\gamma_{k,2}^{\left(o\right)}\right)$ where $\gamma_{k,2}^{\left(o\right)}$ is calculated with~\eqref{DelayAware08} and $\overline{\mathbf{w}}_{g}^{\left(o\right)}$, $\forall g\in\mathcal{G}$. In~\eqref{DelayAware34d}, $\mathcal{V}^{\left(o\right)}\triangleq\left\{\mathbf{P}_{g,i}^{T}\left(1\right)
\overline{\mathbf{w}}_{g}^{\left(o\right)}\right\}_{g\in \mathcal{G},i\in\mathcal{K}_{\mathrm{R}}}$ and $\mathcal{O}^{\left(o\right)}\triangleq\left\{\mathbf{\Omega}_{i}^{\left(o\right)}\right\}_{i\in\mathcal{K}_{\mathrm{R}}}$.

\subsection*{D. Initialization of Algorithm~\ref{DelayAwareAlg02} and~\ref{DelayAwareAlg03}}
In Algorithm~\ref{DelayAwareAlg02} and~\ref{DelayAwareAlg03}, the initialization of $\mathbf{w}_{g,i}$, $\mathbf{u}_{g,i}$, $\mathbf{v}_{g,i}$, $\forall g\in\mathcal{G}$ and $\mathbf{\Omega}_{i}$, $\forall i\in\mathcal{K}_{\mathrm{R}}$ is finished via two steps. First, the initial values of $\mathbf{w}_{g,i}$, $\mathbf{u}_{g,i}$, $\forall g\in\mathcal{G}$ are randomly chosen and the initial values of $\mathbf{v}_{g,i}$, $\forall g\in\mathcal{G}$ and $\mathbf{\Omega}_{i}$, $\forall i\in\mathcal{K}_{\mathrm{R}}$ are given by:
\begin{subequations}\label{DelayAware47}
\begin{align}
&\mathbf{v}_{g,i}=
\begin{cases}
\sqrt{\delta\frac{e^{C_{i}}-1}{\sum\limits_{g\in\mathcal{G}}
\overline{c}_{f_{g},i}}}\mathbf{v}_{i}, & \sum\limits_{g\in\mathcal{G}}
\overline{c}_{f_{g},i}\neq 0 \\
\mathbf{0}, & \sum\limits_{g\in\mathcal{G}}
\overline{c}_{f_{g},i}= 0
\end{cases}\label{DelayAware47a}\\
&\mathbf{\Omega}_{i}=\mathbf{I}_{N_{t}} \label{DelayAware47b}
\end{align}
\end{subequations}
where $\mathbf{v}_{i}$ is a normalized random vector and $0<\delta\leq 1$. Then, they are normalized such that power constraint~\eqref{DelayAware35d} is satisfied. It is easy to observe that fronthaul capacity constraint~\eqref{DelayAware20e} is satisfied when the values of $\mathbf{v}_{g,i}$, $\forall g\in\mathcal{G}$ and $\mathbf{\Omega}_{i}$, $\forall i\in\mathcal{K}_{\mathrm{R}}$, are given by~\eqref{DelayAware47}.

\subsection*{E. Convexity of Problem~\eqref{DelayAware35}}

In this subsection, we focus on convexifying problem~\eqref{DelayAware35} via the PPD method and SCA method. Similarly, constraint~\eqref{DelayAware35c} can be approximated by
\begin{equation}\label{DelayAware37}
\mathrm{r}_{g,1}-\ln\left(\mu_{k,1}\right)
+\phi\left(\chi_{k,1},\chi_{k,1}^{\left(t\right)}\right)\leq 0, \forall k\in\mathcal{K}_{\mathrm{U}}.
\end{equation}
Now, we turn our attention to transform objective~\eqref{DelayAware35a} and constraint~\eqref{DelayAware35e} into a convex form. Introducing auxiliary variables $\eta$ and $\theta$, problem~\eqref{DelayAware35} can be equivalently rewritten as
\begin{subequations}\label{DelayAware38}
\begin{align}
&\min~\eta+\theta\label{DelayAware38a}\\
\mathrm{s.t.}~&\eqref{DelayAware20c}, \eqref{DelayAware20d},\eqref{DelayAware23}, \eqref{DelayAware24},\eqref{DelayAware35d}, \eqref{DelayAware37}\label{DelayAware38b}\\
&\left(\tau_{0}+\theta\right)\mathrm{r}_{g,1}\leq S, \forall g\in\mathcal{G}, \label{DelayAware38c}\\
&S-\tau\mathrm{r}_{g,1}\leq\eta \mathrm{r}_{g,2}\label{DelayAware38d}\\
&\theta=\frac{S}{\min\limits_{i\in\mathcal{K}_{\mathrm{R}}}\left(\ln\left(\left|\mathbf{A}_{i}
\right|\right)-\ln\left(\left|\mathbf{\Omega}_{i}\right|\right)\right)}\label{DelayAware38e}
\end{align}
\end{subequations}
where the optimization variables are $\eta$, $\theta$, $\mathbf{W}_{g}$, $\overline{\mathbf{W}}_{g}$, $\mathbf{\Omega}$, $\mathrm{r}_{g,p}$, $\forall g\in\mathcal{G}$, $\forall p\in\mathcal{P}$. Note that constant $\tau_{0}$ in the objective function~\eqref{DelayAware38} is omitted. Further, introducing auxiliary $\kappa_{g}$, $\psi_{g}$, $\forall g\in\mathcal{G}$, after some basic mathematical operation, problem~\eqref{DelayAware38} is equivalently reformulated as
\begin{subequations}\label{DelayAware39}
\begin{align}
&\min~\eta+\theta\label{DelayAware39a}\\
\mathrm{s.t.}~&\eqref{DelayAware20c}, \eqref{DelayAware20d},\eqref{DelayAware23}, \eqref{DelayAware24},\eqref{DelayAware35d}, \eqref{DelayAware37}\label{DelayAware39b}\\
&\left(\tau_{0}+\theta\right)\mathrm{r}_{g,1}\leq S, \forall g\in\mathcal{G} \label{DelayAware39c}\\
&\psi_{g}\leq\theta\mathrm{r}_{g,1}, \forall g\in\mathcal{G}\label{DelayAware39d}\\
&\kappa_{g}\leq\eta\mathrm{r}_{g,2},\forall g\in\mathcal{G}\label{DelayAware39e}\\
&S-\tau_{0}\mathrm{r}_{g,1}-\psi_{g}-\kappa_{g}\leq 0, \forall g\in\mathcal{G}\label{DelayAware39f}\\
&\theta=\frac{S}{\min\limits_{i\in\mathcal{K}_{\mathrm{R}}}\left(\ln\left(\left|\mathbf{A}_{i}
\right|\right)-\ln\left(\left|\mathbf{\Omega}_{i}\right|\right)\right)}\label{DelayAware39g}
\end{align}
\end{subequations}
where the optimization variables are $\eta$, $\theta$, $\kappa_{g}$, $\psi_{g}$, $\mathbf{W}_{g}$, $\overline{\mathbf{W}}_{g}$, $\mathbf{\Omega}$, $\mathrm{r}_{g,p}$, $\forall g\in\mathcal{G}$, $\forall p\in\mathcal{P}$. At the optimal point of problem~\eqref{DelayAware39}, constraints~\eqref{DelayAware39d},~\eqref{DelayAware39e}, and~\eqref{DelayAware39f} are activated. Using the monotonic property of function $\ln\left(\cdot\right)$, constraints~\eqref{DelayAware39c},~\eqref{DelayAware39d}, and~\eqref{DelayAware39e} can be transformed into a difference of convex functions form, i.e.,
\begin{subequations}\label{DelayAware40}
\begin{align}
&\ln\left(\tau_{0}+\theta\right)+\ln\left(\mathrm{r}_{g,1}\right)-\ln\left(S\right)\leq 0\label{DelayAware40a}\\
&\ln\left(\psi_{g}\right)-\ln\left(\theta\right)-\ln\left(\mathrm{r}_{g,1}\right)\leq 0, \forall g\in\mathcal{G}\label{DelayAware40b}\\
&\ln\left(\kappa_{g}\right)-\ln\left(\eta\right)-\ln\left(\mathrm{r}_{g,2}\right)\leq 0. \forall g\in\mathcal{G}\label{DelayAware40c}
\end{align}
\end{subequations}
Further, exploiting the concavity of function $\ln\left(\cdot\right)$,~\eqref{DelayAware40} can be approximated to the following convex form
\begin{subequations}\label{DelayAware41}
\begin{align}
&\phi\left(\tau_{0}+\theta,\tau_{0}+\theta^{\left(t\right)}\right)+\phi\left(\mathrm{r}_{g,1},
\mathrm{r}_{g,1}^{\left(t\right)}\right)-\ln\left(S\right)\leq 0\label{DelayAware41a}\\
&\phi\left(\psi_{g},\psi_{g}^{\left(t\right)}\right)-\ln\left(\theta\right)-\ln\left(\mathrm{r}_{g,1}\right)\leq 0, \forall g\in\mathcal{G}\label{DelayAware41b}\\
&\phi\left(\kappa_{g},\kappa_{g}^{\left(t\right)}\right)-\ln\left(\eta\right)-\ln\left(\mathrm{r}_{g,2}\right)\leq 0. \forall g\in\mathcal{G}.\label{DelayAware41c}
\end{align}
\end{subequations}
Replacing constraints~\eqref{DelayAware39c},~\eqref{DelayAware39d}, and~\eqref{DelayAware39e} with inequalities~\eqref{DelayAware41a},~\eqref{DelayAware41b}, and~\eqref{DelayAware41c}, respectively, problem~\eqref{DelayAware39} can be further approximated to
\begin{subequations}\label{DelayAware42}
\begin{align}
&\min~\eta+\theta\label{DelayAware42a}\\
\mathrm{s.t.}~&\eqref{DelayAware39b}, \eqref{DelayAware39f}, \eqref{DelayAware41} \label{DelayAware42b}\\
&\theta=\frac{S}{\min\limits_{i\in\mathcal{K}_{\mathrm{R}}}\left(\ln\left(\left|\mathbf{A}_{i}
\right|\right)-\ln\left(\left|\mathbf{\Omega}_{i}\right|\right)\right)}\label{DelayAware42c}
\end{align}
\end{subequations}
where the optimization variables are $\eta$, $\theta$, $\kappa_{g}$, $\psi_{g}$, $\mathbf{W}_{g}$, $\overline{\mathbf{W}}_{g}$, $\mathbf{\Omega}$, $\mathrm{r}_{g,p}$, $\forall g\in\mathcal{G}$, $\forall p\in\mathcal{P}$. Similar to problem~\eqref{DelayAware26}, the main obstacle of solving problem~\eqref{DelayAware42} is equality constraint~\eqref{DelayAware42c}.

In the following, we resort to the penalty dual decomposition method to solve problem~\eqref{DelayAware42}. The partial augmented Lagrangian function of problem~\eqref{DelayAware42} is given by
\begin{subequations}\label{DelayAware43}
\begin{align}
&\min~\eta+\theta+\frac{1}{2\rho}\left|\frac{S}{\theta}-\min\limits_{i\in\mathcal{K}_{\mathrm{R}}}\left(\ln\left(\left|\mathbf{A}_{i}
\right|\right)-\ln\left(\left|\mathbf{\Omega}_{i}\right|\right)\right)
+\rho\lambda\right|^{2}\label{DelayAware43a}\\
\mathrm{s.t.}~&\eqref{DelayAware39b}, \eqref{DelayAware39f}, \eqref{DelayAware41}\label{DelayAware42b}
\end{align}
\end{subequations}
where the optimization variables are $\eta$, $\theta$, $\kappa_{g}$, $\psi_{g}$, $\mathbf{W}_{g}$, $\overline{\mathbf{W}}_{g}$, $\mathbf{\Omega}$, $\mathrm{r}_{g,p}$, $\forall g\in\mathcal{G}$, $\forall p\in\mathcal{P}$. In problem~\eqref{DelayAware43}, $\lambda$ is the Lagrange multiplier and $\rho$ is a scalar penalty parameter. Following a procedure similar to that used for problem~\eqref{DelayAware27}, to further obtain a tractable form of problem~\eqref{DelayAware43}, we relax problem~\eqref{DelayAware43} to the following form
\begin{subequations}\label{DelayAware44}
\begin{align}
&\min~\eta+\theta+\frac{1}{2\rho}\sum\limits_{i\in\mathcal{K}_{\mathrm{R}}}
\mathds{1}\left(\sum\limits_{g\in\mathcal{G}}\overline{c}_{f_{g},i}\right)\left|\frac{S}{\theta}+
\ln\left(\left|\mathbf{\Omega}_{i}\right|\right)-\ln\left(\left|\mathbf{A}_{i}
\right|\right)+\rho\lambda\right|^{2}\label{DelayAware44a}\\
\mathrm{s.t.}~&\eqref{DelayAware39b}, \eqref{DelayAware39f}, \eqref{DelayAware41}\label{DelayAware44b}
\end{align}
\end{subequations}
where the optimization variables are $\eta$, $\theta$, $\kappa_{g}$, $\psi_{g}$, $\mathbf{W}_{g}$, $\overline{\mathbf{W}}_{g}$, $\mathbf{\Omega}$, $\mathrm{r}_{g,p}$, $\forall g\in\mathcal{G}$, $\forall p\in\mathcal{P}$. Further, using the SCA method to convexify objective~\eqref{DelayAware44a}, problem~\eqref{DelayAware44} can be approximated as the following convex upper bound problem
\begin{subequations}\label{DelayAware45}
\begin{align}
&\min~\eta+\theta+\frac{1}{2\rho}\sum\limits_{i\in\mathcal{K}_{\mathrm{R}}}
\mathds{1}\left(\sum\limits_{g\in\mathcal{G}}\overline{c}_{f_{g},i}\right)\left|\frac{S}{\theta}+
\phi\left(\mathbf{\Omega}_{i},\mathbf{\Omega}_{i}^{\left(t\right)}\right)-\ln\left(\left|\mathbf{A}_{i}
\right|\right)+\rho\lambda\right|^{2}\label{DelayAware45a}\\
\mathrm{s.t.}~&\eqref{DelayAware39b}, \eqref{DelayAware39f}, \eqref{DelayAware41}\label{DelayAware45b}
\end{align}
\end{subequations}
where the optimization variables are $\eta$, $\theta$, $\kappa_{g}$, $\psi_{g}$, $\mathbf{W}_{g}$, $\overline{\mathbf{W}}_{g}$, $\mathbf{\Omega}$, $\mathrm{r}_{g,p}$, $\forall g\in\mathcal{G}$, $\forall p\in\mathcal{P}$. 

\begin{small}

\end{small}
\end{document}